\documentclass[12pt,preprint]{aastex}
\def\kms    {\ifmmode{{\rm ~km~s}^{-1}}\else{~km~s$^{-1}$}\fi}
\def\lsun   {\ifmmode{{\rm ~L}_\odot}\else{~L$_\odot$}\fi}
\def\msun   {\ifmmode{{\rm ~M}_\odot}\else{~L$_\odot$}\fi}
\shorttitle{Water Masers in NGC 2071}
\shortauthors{Seth, Greenhill, \& Holder}

\begin{document}

\slugcomment{Resubmitted 09-Aug-02} 

\title{ Water Masers as Tracers of Protostellar Disks and Outflows in the 
Intermediate Mass Star Forming Region NGC\,2071}

\author{A. C. Seth}
\affil{University of Washington}
\email{seth@astro.washington.edu}

\author{L. J. Greenhill}
\affil{Harvard-Smithsonian Center for Astrophysics}
\email{greenhill@cfa.harvard.edu}
\and
\author{B. P. Holder}
\affil{University of Texas}
\email{bholder@physics.utexas.edu}

\begin{abstract}

We have mapped the water maser emission associated with the infrared centers
IRS1 and IRS3 of the NGC\,2071IR star forming region at four epochs over
$\sim 4$ months with the Very Long Baseline Array (VLBA).  We detected 269
maser features with $\sim 1$\kms~linewidths and measured 30 proper motions. In
each infrared center, the water maser emission appears to trace parts of a
protostellar disk and collimated outflow. The disk components are $\sim 9$ and
$\sim 17$ AU long, in IRS\,3 and IRS\,1 respectively, and $\sim 2$ AU wide.
They are identified as disks by their compact size, elongation parallel to the
direction of known IR polarization, central location in the maser maps, small
internal proper motions, and proximity to $\lambda1.3$\,cm continuum emission.
The outflows have axes perpendicular to the disks and exhibit proper motions
of up to $\sim 42$\kms. They are outlined by maser emission up to $\sim 260$
AU from the protostars. The IRS\,3 outflow appears to be conical on one side,
while the IRS\,1 outflow comprises a  narrowly collimated bipolar flow
surrounded by outward-facing, funnel-shaped cavities. The detection of water
maser emission tracing such compact disk components and specifically conical or
funnel-shaped structures is unusual.  The fact that the distributions are
similar in IRS\,3 and IRS\,1 may indicate the two infrared centers are roughly
coeval.  NGC\,2071IR provides a rare opportunity to resolve the structures and
dynamics of disks and outflows together, and to do so for two protostars that
are only $\sim 2000$ AU apart (in projection) in a deeply embedded star
forming region of intermediate luminosity.

\end{abstract}


\section{Introduction}

The infrared cluster NGC\,2071IR lies in the Lynds 1630 dark cloud in Orion at
a distance of 390 pc \citep{anthony-twarog82}.  The $\sim 30''$ diameter
cluster has been resolved into 8 distinct near infrared sources
\citep{walther93} and has a total luminosity of
520\,L$_{\odot}$
\citep{butner90}, which is suggestive of intermediate mass star formation. 
\citet{snell86} first detected radio continuum counterparts to the IR sources,
specifically IRS1 and IRS3 separated by 6''. While IRS\,1 dominates the
luminosity at near infrared  wavelengths, IRS\,3 is a significant contributor
at longer wavelengths \citep{snell86,kawabe89}.  

NGC\,2071IR also hosts a well-studied molecular outflow that has been mapped in
the emission of CO \citep{scoville86, moriarty-schieven89, chernin92}, H$_2$
\citep{garden90,aspin92}, CS \citep{zhou91,kitamura92}, SO and SiO 
\citep{chernin93}, HCO+ \citep{girart99} and  NH$_{3}$ \citep{zhou90}. 
The arcsecond-resolution H$_2$ observations of \citet{aspin92} indicate
IRS\,1 is the likely source for the large scale outflow while those of 
\citet{garden90} show elongated molecular emission associated with IRS\,3 as well.
Images of radio continuum emission,
with resolutions as high as $0\rlap{.}''1$, show elongated emission from 
thermal jets coincident with both infrared sources \citep{torrelles98,smith94,snell86}.

Water maser emission ultimately associated with NGC\,2071IR was detected even
before the IR cluster and molecular outflows
\citep{schwartz75,pankonin77,campbell78}.  Both IRS\,1 and IRS\,3 host the
water maser emission \citep{tofani95,torrelles98}, which indicates the
presence of substantial columns of dust laden, warm (300 - 1000 K), dense
($10^8$ - $10^{10}$ cm$^{-3}$) gas that is probably shock excited
\citep{elitzur92}.  \citet{torrelles98} observed the masers  using the VLA
with resolution of $0\rlap{.}''1$.  In IRS\,1, they observed a distribution of
masers elongated parallel to the roughly east-west radio jet.  In IRS\,3, they
observed a distribution more or less perpendicular to the jet and suggestive
of a disk.  The difference in the structure for the maser sources,
supplemented by the higher extinction toward IRS\,3, was inferred to indicate
IRS\,3 is less evolved. 

In general, the study of intermediate and high-mass protostars is difficult
because examples are hundreds of parsecs away and are contained within crowded
fields.  As a result, even observations with $0\rlap{.}''1$ resolution can be
confusion limited. However, long baseline interferometric observations, with
milliarcsecond (mas) resolution,  can be used to map regions unambiguously (and
estimate proper motions) when high brightness temperature emission, such as
water maser emission, is present \citep[e.g.,][and references therein]{claussen98,moscadelli00,patel00,furuya00,torrelles01,gwinn92}.

The proper motion study of NGC\,2071IR presented here concentrates on one of
the closest regions of intermediate mass star formation whose
structure and dynamics we show have not been fully resolved in previous
studies.  In Section 2, we present the observations and calibration, followed
by a  description of how we estimated proper motions in Section 3.  In Section
4 we discuss the compact protostellar disk and outflows found in both IRS\,1 and
IRS\,3, and consider how these structures bear upon other lower angular resolution 
observations of the whole infrared cluster.  
Conclusions are presented in Section 5.

\section{Observations and Calibration}

We observed the $6_{16}-5_{23}$ transition of water (rest frequency 22235.080
MHz) toward NGC\,2071IR with the Very Long Baseline Array (VLBA) of the
NRAO\footnote{The National Radio Astronomy Observatory is a facility of the
U.S. National Science Foundation operated under cooperative agreement by
Associated Universities, Inc.} at four epochs in 1996 (see Table\,1). A single
antenna of the Very Large Array (VLA) augmented the VLBA, which provided
improved sensitivity to emission extended on several mas scales. A 16 MHz
($\sim215$\kms) recording bandwidth, centered on
$V_{\rm LSR}=4.66$\kms~covered the known velocity range of emission. We
processed the data with the VLBA correlator, obtaining 512 spectral
channels with a separation of $\sim 0.42$\kms.

The data were calibrated and images constructed using the NRAO AIPS package and
standard techniques for VLBI.   All four data sets were  shifted to a common
phase center located at the radio continuum peak of IRS\,1 \citep{torrelles98}
after correlation.  We set the flux density scale  with 30\% uncertainty by
template fitting total-power spectra for each antenna, where the template
spectrum was itself calibrated with system temperature and published gain
curves.  We calibrated interferometric delay and bandpass response using
observations of 0234+285, 0528+134, and NRAO150.  To remove the effects of
atmospheric pathlength fluctuations, we self-calibrated each epoch using the
strong, isolated emission at $V_{\rm LSR}=12.75$\kms, which originates in
IRS1. We applied this self-calibration to data for all spectral channels,
created synthesis images of the IRS\,1 and IRS\,3 fields simultaneously, and
detected emission between $-2.42$ and $27.92$\kms~in IRS\,1 and $-8.69$ and
$27.12$\kms~in IRS\,3. No emission was detected in association with other
infrared centers in NGC\,2071IR, consistent with \citet{torrelles98} and
\citet{tofani95}.  The calibration error budget is dominated by a 16 cm
uncertainty in the VLA station position (estimated with respect to a USNO
geodetic reference frame). This causes a $<10$ microarcsecond
($\mu$as)  position uncertainty for
masers in IRS\,3, which is small compared to the month-to-month $\sim
220 \mu$as change that corresponds to a modest 5\kms~proper motions.

\section{Identification of Maser Features and Proper Motions}

We identified all emission components stronger than $6\times$ the RMS noise
level in the images for each spectral channel. We fit one or more model
elliptical Gaussians as needed to decompose these components into individual
maser ``spots'' each about the same size as the interferometer beam.  
Position uncertainties for the spots were typically 0.05 mas.  Counting all
four epochs, 1074 maser spots were detected, with $\sim 40\%$ in IRS\,3
(Table\,1).

We grouped the maser spots into emission features with resolved line
profiles. Each feature probably represents a distinct clump of gas.  Over
the four epochs, we detected 269 features (Table\,1). To group the
spots into features, we selected spots in
adjacent channels that lay within one beamwidth of each other.  In
ambiguous cases (e.g., a close pair of maser spots in a single channel)
we selected the closest spot.  For each feature, we estimated the
position from the flux-weighted mean position of the contributing maser
spots. The uncertainty was the greater of the formal error in the mean
and the standard deviation of the spot positions.

For each maser feature detected, we also estimated spectral
characteristics. We fit Gaussian spectral-line profiles to  centrally peaked
features comprising three or more channels.  For other features we simply
adopted the flux-weighted mean velocity and maximum flux density among the
contributing channels. Because the ({\it u,v})-coverage of the VLBA is flattened
and sparse for sources close to the celestial equator, artifacts were common in
a north-south band around bright emission. Features that lay in these bands
were discounted unless they were offset in velocity.  The final tally of maser
features in each channel accounts typically for 50 to 100\% of the flux density
detected in a total-power spectrum of the NGC\,2071IR region (Figure\,1), which
suggests that most of the maser features are small compared to the $\sim 1$ mas
(0.39 AU) VLBA beam, and that the correct identification of imaging artifacts
was largely successful.  We present the maser feature line-of-sight 
velocities, positions, and other characteristics for epochs 1 to 4 in Table\,2.

After identifying features, we registered our VLBA map and the VLA map of
\citet{torrelles98}, which corresponded to an epoch $\sim 4$ months after the
close of the VLBA observations.  Using three persistent clumps of maser
emission in IRS\,1 as reference points, we aligned the maps with an estimated
uncertainty of $0\rlap{.}''01$ (corresponding primarily to uncertainty in the
relative positions estimated with the VLA and proper motions between the VLBA
and VLA epochs). This registration enabled us to compare our maser positions
in both IRS\,1 and IRS\,3 with the maser and continuum positions of
\citet{torrelles98}.

We estimated proper motions via weighted linear least squares fits for (VLBA)
features that were (1) present in at least three epochs, (2) stationary to
$<0.5$\kms~along the line-of-sight (LOS), and (3) moving in a straight line at
constant velocity.   To do this, first, the images for the four epochs were
registered by aligning the feature at V$_{\rm LSR}=8.76$\kms~in IRS\,1 and
then the 3.15\kms~feature in IRS\,3.  The separate registration of features
within IRS\,3 was necessary due to a $\sim3$ mas offset for the first epoch;
feature positions for the other epochs were not offset significantly. Second,
for features in IRS\,1, the mean proper motion was subtracted to compensate
for possible motion of the 12.75\kms~reference feature. Because positions and
motions are measured in a relative sense, any plausible transverse velocity
may be subtracted. Subtracting the mean proper motion is the same as assuming
that IRS\,1 is stationary (as sampled by the available motions).   Due to the
small number of motions in IRS\,3, it was not possible to obtain a meaningful
mean proper motion.  The IRS\,3 proper motions therefore include the motion of
the 3.15\kms~reference feature located in the disk. The mean formal error in
the magnitude and direction of all detected motions were $9\%$ and $5^\circ$,
respectively.  The medians were $3\%$ and $1^\circ$, respectively.  We present
the proper motions for 30 maser features in IRS\,3 and IRS\,1 in Table\,3.

\section{Discussion}

The distributions of water maser emission in IRS\,3 (Figure\,2) and IRS\,1
(Figure\,3) are complex and difficult to model with absolute certainty. We
suggest relatively simple interpretations, in the context of the current star
formation paradigm. In each infrared source, the masers trace a compact
protostellar disk and a collimated outflow.  We suggest this model because
each source contains
\begin{enumerate}
\item a centrally located, elongated clump of maser
features $\sim 9$ AU (IRS\,3) or $\sim 17$ AU (IRS\,1) long, and $\sim 2$ AU
wide, with small internal proper motions, orientation parallel to the
known direction of local IR polarization
\citep{walther93}, and in close proximity to $\lambda1.3$ cm continuum peaks
\citep{torrelles98}; and
\item maser features up to $\sim 260$ AU from the putative disk in which
flow is oriented away from the disk and with proper motions of 17 to
42\kms.
\end{enumerate}
We refer to these two populations as the ``disk masers'' and the ``peripheral
masers,'' respectively.  Figure\,4 outlines the main elements of the models
for IRS\,3 and IRS\,1.

\subsection{IRS 3}

\subsubsection{The Disk}

The central clump of water maser features in IRS\,3, which we propose traces
the near side of a roughly edge-on disk, is $\sim 9\times2$ AU in size at a
position angle (PA) of $\sim -55^\circ$ (Figure\,2b).   This PA is consistent
with the $-66^\circ$ PA for $2 \mu$m polarization observed toward IRS\,3 by
\citet{walther93}, which is suggestive of scattering by material above and
below the plane of an obscuring disk. In addition, the clump lies within the
$1\sigma$ error circle ($\pm 10$ mas) of the $\lambda 1.3$ cm emission peak of
the compact radio jet (PA$\sim 11^\circ$) mapped by \citet{torrelles98}, which
is suggestive of a physical relationship between the two.

However, no compact distribution of disk masers was detected by
\citet{torrelles98}.  This is probably because the observations were limited
by confusion. Nearly all of the IRS\,3 maser features lie within $\sim 1$
VLA beamwidth, $\sim 0\rlap{.}''1$ (Figure\,2a).  Because the disk masers are
not strong they could easily be blended with stronger masers in the nearby
outflow.  We note as well that over the $\sim 4$ months we observed, the
maximum flux density of the disk masers decreased with time (Table\,2).
Since the Torrelles et al. observations trailed our own by another $\sim 4$
months, the effects of blending may have been accentuated. Nonetheless, we
note that although a rich population of disk masers was not visible, Torrelles
et al. did detect four masers with velocities close to the LOS velocity of the
putative disk ($\sim 4$\kms) and within one VLA beamwidth of its location
(Figure\,2).

Surprisingly, we do not detect among the disk masers a systematic velocity
gradient characteristic of rotation.  The mean LOS velocity of the disk masers
is $4.1$\kms~with a range of 3.8\kms (Table\,2).  Hence, any
gradient is $<4$\kms~over 9 AU (i.e., the total range in the Doppler
velocities),  which is suggestive of an improbably small encircled mass, 
$\ll 0.1$ M$_\odot$.  We conclude that the maser features lie at radii $>9$
AU, where  the orbital speed may be expected to be small, and that the masers
sample only a confined sector on the front side of the disk.   More
specifically, rotating edge-on disks display gradients in LOS velocity,
${{\partial V_{los}} / {\partial b}}= 30\sqrt{M/r^3}$, where $V_{los}$ is
line-of-sight velocity in\,\kms, $b$ is impact parameter with respect to the
center of the disk measured in AU, $M$ is enclosed mass in M$_\odot$, and $r$
is disk radius in AU. For $r=4.5$~AU and $M=1$~M$_\odot$, the observed LOS
velocity should change by $\sim 28$\kms~across the IRS\,3 clump, a possibility
the data exclude.  However, if the masers sample a small sector on the near
side of a larger disk, then the upper limit on the observed gradient, a
$<4$\kms~shift over $\sim 9$ AU, constrains the radius of the maser region in
the disk to be
$>17M^{1/3}$ AU.   Because the excitation of maser
emission requires a balance of physical conditions (i.e., temperature and
density), which probably exists only over a narrow range of radius, the disk
itself could extend to larger or smaller radii.  Maser action also depends on
the magnitude of gradients in LOS velocity along the LOS through the disk. 
These gradients reach a minimum in a small sector along the front side of a
disk \citep[e.g.,][]{ponomarev94}, which is consistent with the suggested
placement of maser features.

The disk masers are blueshifted by $\sim 6$\kms (Figure\,2b) with respect to
the 9.7\kms~systemic velocity for NGC\,2071IR obtained from CO and HCO$^+$
observations \citep{moriarty-schieven89, girart99}.  (We will assume the
systemic velocity for both IRS\,1 and IRS\,3 is 9.7\kms.)   This Doppler shift
is suggestive of expansion, where either (1) the disk is expanding and is warm
(and dense) enough to support widespread maser action, or (2) the
warm maser-laden surface layers of a cool, optically thick disk, are swept
outward by oblique wind-disk collision or by protostellar radiation, as could
occur during photo-evaporation \citep[e.g.,][]{hollenbach94}.  The net
blueshift of the IRS\,3 disk masers is larger than their proper motion
(1.8\kms) with respect to IRS\,1.  This lends credibility to the suggestion
that the blueshift of the IRS\,3 disk masers is internal to IRS\,3, e.g.,
expansion.  (To measure the 1.8\kms~proper motion, we subtracted the mean
motion of IRS\,1 (see Section\,3) but estimated proper motions in IRS\,3
without aligning the maps on the 3.15\kms~reference feature. We also used only
the last three epochs.)

We note that a small clump of features with a 6\kms~redshift (relative to
systemic) lies 14 mas northeast of the blueshifted  disk features (Figure\,2b).
We speculate that these could mark outflowing material on the far side of the
disk.  If in fact we have resolved the separation between approaching (front)
and receding (back) sections of the disk, then we can constrain the disk
radius (as before) and inclination together.  For a 1 M$_{\odot}$ protostar,
the radius of the maser region in the disk is  $>16$ AU, and it is tilted
$<10^{\circ}$ from edge-on.  Greater mass entails larger radii and smaller
tilts.  We note that in principle these blue and redshifted maser features
could also signify infall, but we favor outflow because we expect the more
prominent  maser emission to be on the nearside of the disk, and we have
inferred that the axis of the bipolar outflow is inclined slightly toward us
and to the east (see next section).  

\subsubsection{The Conical Outflow}

The peripheral masers are distributed irregularly over a
$230\times270$ AU region (Figure\,2a) and exhibit proper motions of 22 to
42\kms~(Table\,3), which are primarily directed away from the disk. These
proper motions are large compared to the 5\kms~mean proper motion within the
disk (Table\,3). The magnitude and direction of these masers are strongly
suggestive of an outflow. The mean LOS velocity of the peripheral masers is
11\kms, in reasonable agreement with the 9.7\kms~systemic velocity.  The
outflow to the north and east of the disk has a mean velocity of
$\sim8$\kms~and is thus slightly blueshifted,  while the much
more sparsely traced outflow south and west of the disk is redshifted, with a
mean velocity of $\sim19$\kms~(Table\,2).

The distribution of the outflowing
masers to the north and east is similar in appearance to the limbs of an
outflow cone 20 to 90 AU in length, which is seen most clearly in Figure\,2b,
but also includes the northernmost blue-shifted masers visible in Figure\,2a.
The cone has a $\sim 110^\circ$ opening angle centered on the disk minor axis
(Figure\,4a). Based on relative LOS velocities and proper motions it is
probably inclined  on the order of $10^\circ$ with respect to the plane of the
sky, although uncertainty in the systemic velocity of IRS\,3, and sparse
sampling of the outflow make the inclination difficult to estimate.
The limbs of the proposed outflow cone could mark an interface between
infalling and outflowing material, a scenario discussed theoretically by
\citet{delamarter00}. Such an interface would be a natural site of maser
emission because of energy deposited locally by shocks \citep{elitzur92}. If
the boundary layer is thin, the long velocity-coherent path lengths necessary
for maser amplification will lie along the edges of the cone, close to the
plane of the sky, giving the impression of limb-brightening.  We note a $\sim
11$\kms~difference in Doppler velocity between the two limbs of the cone.  This could be attributed to
the angular momentum of the infalling material.  A rotation velocity on the
order of 5\kms~is roughly consistent with an enclosed mass of 1 M$_\odot$
and radius of 50 AU.  The sense of the Doppler shift (blueshift along the
northern limb) is the same as the direction of the velocity gradient detected
by \citet{torrelles98}, which they interpreted as a manifestation of disk
rotation, but which we suggest is dominated by the velocity structure
of the peripheral masers.

The outflow cone can be traced to within $\sim 20$ AU of the disk (Figure\,2b),
which suggests that collimation occurs at or within that radius.  On the
reverse side of the disk, the sparser appearance of the red-shifted outflow is
difficult to explain with certainty. Because maser action requires specific
physical and dynamic conditions, the absence of emission does not necessarily
imply the absence of warm dense gas, as in an outflow cone. As well, the
collimating medium could be asymmetrically distributed about the disk plane,
discouraging formation of a sharp cone on one side. Alternatively, a
collimated redshifted outflow could exist but be obscured by gas ionized and
driven outward by the IRS\,3 protostar as the surrounding interstellar medium
is heated. Such an uncollimated expansion could be consistent with the two
proper motions most widely separated from the disk. 

\subsection{IRS 1}

\subsubsection{The Disk}

The central clump of water masers in IRS\,1, which we propose traces the front
side of a protostellar disk, is $\sim 17$ AU long and $\sim 2$ AU wide, at a PA
of $\sim -11^\circ$ (Figure\,3b).  This PA is similar to the $-19^\circ$ PA of
the 2$\mu$m polarization vectors observed by \citet{walther93}, who suggest
that the high polarization may be related to an obscuring disk. The peak of the
compact radio continuum jet mapped by \citet{torrelles98} lies $\sim 35$ mas
to the east and along the minor axis of the disk.  Overall, the distributions
of masers detected with the VLA and VLBA agree well (Figure\,3a), probably
because IRS\,1 is extended over many VLA beamwidths, which reduces the impact
of the blending of maser features.  However, as in IRS\,3, only the VLBA
detects a compact disk component. \citet{torrelles98} detected only one maser
in close proximity to the putative disk, about which we cannot hypothesize.

The disk masers exhibit Doppler velocities of 6.0 to 19\kms, with the
blueshifted masers lying to the north and the redshifted masers to the south 
(Figure\,3b; Table\,2).  This velocity gradient may be explicit evidence of
rotation.  Following an analysis similar to that applied to the data for
IRS\,3, we infer an enclosed mass of $0.6 (r/10 AU)^3$ M$_\odot$ for an
edge-on geometry.  Independent estimates of mass in IRS\,1 are difficult to
formulate, but given that IRS\,1 is the dominant source of the 520 L$_\odot$
radiated by  NGC\,2071IR
\citep{butner90}, we assume a mass of a  few M$_\odot$. For this mass, the
implied radius of the maser emission is $\sim 20$ AU, which indicates the
observed maser emission traces only a limited sector on the nearside of the
disk.  We note that the mean velocity of the disk is 13\kms~(Table\,2), a
$\sim 3$\kms~redshift from our assumed systemic velocity.  Just as the
blueshift of maser features in IRS\,3 might suggest outflow coupled with
rotation, this redshift in IRS\,1 may imply that the material being traced by
the masers is infalling.  However, we treat the possible detection of
infall tentatively because the redshift is small and the proper motions
of disk masers (see below) are somewhat disordered, given the simple model we
propose.

We have detected three proper motions in the disk, two of which have
significant southward velocity components of 5 and 21\kms (Figure\,3b;
Table\,3). The expected orbital motion is the product of the observed velocity
gradient and radius, independent of enclosed mass, and may be expressed as
$8 (r/10AU)$\kms~towards the south.  For a radius of 20 AU, we would predict a
proper motion of 16\kms, which is consistent with the observed motions. 
However, there is significant scatter both in the direction and magnitude of
the observed motions, and better data are needed to support a complete
analysis.  The proper motions could be  influenced by turbulence, radial
flows, changing orbital velocity along the line of sight, and inclination
effects.  Relatively few motions were measured in the disk because the
lifetimes of many maser features therein were shorter than the length of the
proper motion study.  In contrast, features in the surrounding outflow were
longer lived.  This observation should be used to guide the design of
future proper motion studies.  

\subsubsection{The High-Speed Outflow and Surrounding Cavities}

The peripheral masers are distributed over a region $\sim 330\times200$ AU,
elongated more or less east-west (Figure\,3a).  The average Doppler velocity of
these masers is 11\kms, in agreement with the assumed systemic velocity of
9.7\kms.  Based on the distribution and the dynamics of maser features, we
suggest that the peripheral masers trace three separate structures
(Figure\,4b):
\begin{enumerate}
\item a narrow high-speed bipolar outflow oriented perpendicular 
to the disk elongation,
\item four arc-shaped arms extending from the disk that correspond to the
limb-brightened edges of two outward facing, funnel-shaped
cavities along the disk minor axis, and
\item an arc of masers located 275 mas to the west and 100 mas to the south of
the disk (Figure\,3c), which may trace a bow shock.
\end{enumerate}
We discuss the high-speed outflow and cavity in this Section and the arc in the
next Section.

The high-speed outflow appears to be narrowly collimated and oriented along a
PA of $\sim77^\circ$ (Figure\,3a), perpendicular to the disk.  Seven proper
motions were observed in the flow with velocities ranging from 17 to 25\kms.  
The western outflow is clearly blueshifted (mean Doppler velocity of 6\kms), 
while the eastern outflow is redshifted. The motions are inclined from the
plane of the sky by a few degrees closest to the disk, which is consistent
with the assumption that the disk is edge-on.  Further out,  the high-speed
motions are inclined by $\sim 30^\circ$, a result of increasing Doppler shift
with radius.  This may be a signature of precession, however, the most
inclined motions also have significantly rotated position angles suggesting
that the high-speed outflow may be channeled by the surrounding medium.

When we superpose the VLBA and VLA maps of masers in IRS\,1 (Figure\,3a),
we observe four arcs that extend from the disk and appear to outline the limbs
of two outward facing, funnel-shaped  cavities roughly centered along the flow
axis (Figure\,4b).   The association of narrowly collimated outflow  with
wide-angle, possibly wind-blown, cavities is exemplified by some Herbig-Haro
flows \citep[e.g.,][]{reipurth2000} and has been treated theoretically
\citep[e.g.,][]{shu94}.  We suggest that in IRS\,1 the cavity walls represent
an interface between outflowing and infalling material (as discussed earlier
for the outflow cone in IRS\,3), where collimation occurs at radii on the
order of 10 to 50 AU.  

We highlight two systematic patterns among the maser features.   First, the
northern two arcs are blueshifted by a few \kms~with respect to the assumed
systemic velocity while the southern arcs  are redshifted by a few \kms
(Figure\,3a). The asymmetry is similar to that observed in IRS\,3, though it is
smaller in IRS\,1.  It may be attributed to rotation of infalling material,
where speeds of a few \kms~at radii of a few tens to 100 AU is consistent with
an enclosed mass of a few M$_\odot$.  Second, proper motions along the
southern two limbs are directed away from the flow axis and are more or less
orthogonal to the limbs, which is suggestive of an expanding cavity. This is
in agreement with simulations of wide-angle winds by \citet{delamarter00}, and
the evolution of the cavity might bear some similarity to that discussed by
\citet{velusamy98} for the B5 IRS\,1 region.  We note that although most of
the observed proper motions are consistent with an expanding cavity model, two
motions on the northern limbs that are not.  If expansion has in fact been
detected, then the northern motions could be a consequence of localized
turbulence and shock substructure along the cavity wall
\citep[e.g.,][]{gwinn92}.   

\subsubsection{The Arc}

Just to the south of the high-speed outflow, about 275 mas west from the disk
we detect an arc-shaped clump of masers (Figure\,3c). The arc is $\sim 8$ AU
in length, and has 7 to 10\kms~Doppler velocities (Table\,2).  The mean
proper motion of the clump is 3\kms~to the north (Table\,3).   However, while
a number of proper motions have been measured for the clump, they are small 
and the addition of small proper motion offsets, consistent with the
uncertainty in the mean proper motion of IRS\,1, significantly changes their
direction.

Due to its shape and  proximity to the high-speed outflow, we suggest the arc
is a shock,  perhaps a bow shock similar to those traced by water masers and 
observed by \citet{claussen98}, \citet{furuya00} and \citet{torrelles01}.
Shocks are natural sites for water maser emission \citep{elitzur89,kaufman96},
and this one might indicate that the medium around the high-speed outflow is
quite inhomogeneous.  Alternatively, the shock may be located along the near
wall of the wind-blown cavity and its location close to the high-speed outflow
is just a chance superposition.  Another possibility is that the masers are
associated with a second protostellar disk.   At the projected separation of
$\sim 120$ AU from the main disk,  a space velocity of 4\kms~would be
consistent with protostellar masses of $\sim 2$ M$_\odot$.  

The 7 to 10\kms~Doppler velocities of the arc correspond quite well  to the
velocities at which  the VLBA observations resolve-out most of the maser
emission, from  which we infer that the resolved maser emission may be
associated with the arc (Figure\,1).   No other single strucuture detected in
the images corresponds exclusively to this same range of velocity.   The size
of the maser spots in this  region further support this idea - the 72 spots in
the arc are on average 15\% larger than those in IRS\,1 as a whole.  Maser
emission extended along a shock front provides a natural explanation
for these observations.

\subsection{Comparison to Other Observations}

There are a considerable number of observations of 
large-scale outflow and the IR and radio continuum emission from NGC\,2071IR.  
The proposed disk-outflow structure for IRS\,3 and IRS\,1 is supported in part
by (1) the near-infrared polarization observations of
\citet{walther93}, as noted previously, and (2) the morphology of high
angular resolution radio continuum images
\citep{smith94,torrelles98}.  The radio continuum imaging resolves a
roughly east-west structure in IRS\,1, consistent with the PA of the
high-speed flow reported here.  In IRS\,3, the continuum is elongated
at a PA of $11^\circ$, which lies within the reported wide-angle
outflow cone.  \citet{walther93} and \citet{garden90} find a similar
orientation for IRS\,3 in broadband near-infrared light, while for
IRS\,1 the orientation is unclear.

However, the disk-outflow model proposed here for NGC\,2071IR agrees less well
with maps of molecular outflows on scales $>1000$ AU.   H$_2$ observations
conducted by
\citet{aspin92} show a bipolar outflow from  IRS\,1 that seems to dominate the
region.  It is  oriented at PA$\sim 45^\circ$ with blue-shifted emission to
the northeast.   Lower resolution mm-wave observations in CO, CS and SO 
\citep{scoville86,moriarty-schieven89,zhou91,chernin92,chernin93} all show
similar position angles  and orientations.   This is not in good agreement
with the structure of the IRS\,1 water masers.
\citet{aspin92} detect no outflow from IRS\,3, however the maps of
\citet{garden90} clearly show a low level of shocked H$_2$ emission associated
with IRS\,3 at an orientation similar to the IR and radio continuum.

The apparent differences between the outflow structures on large and
small scales may be a consequence of two mechanisms.  
First, the ambient medium may channel the outflows, an effect seen in
other systems \citep[e.g.,][]{reipurth96,bally01}.  The high-speed
water maser flow in IRS\,1 shows some evidence for this, executing a
bend on the order of $30^\circ$.  Second, it is possible that the
outflow observed on large scales is created by a merging of multiple
outflows in NGC\,2071IR, of which the observed flows from IRS\,1 and
IRS\,3 are but two of many.  

\section{Conclusions and Closing Comments}

We have detected 269 water maser features and 30 proper motions around IRS\,3
and IRS\,1 in NGC\,2071IR over four epochs with the VLBA.  In both IR
sources the maser emission appears to trace a protostellar disk and
associated outflow.  The disk components comprise centrally
located, elongated clumps of masers, with position angles similar to those of
local IR polarization and small internal proper motions.  As well, the disk
components lie in close proximity to peaks in the continuum emission from
compact radio jets.  In contrast, the outflow components are characterized by
larger proper motions directed away from the putative disks.  The structures of
the outflows are suggestive of a one-sided conical flow (in IRS\,3) and a
bipolar high-speed flow surrounded by outward-facing, wide-angle cavities (in
IRS\,1).  The limbs of the outflows may mark the interface between outflowing
and rotating infalling material.  

The gross similarity of structure between IRS\,1 and IRS\,3 indicates that
they are at similar stages of development.  In both IR sources, intact
protostellar disks may act as reservoirs of warm dense gas that supports maser
action.  Consequently the inferred disks must be relatively quiescent.  In both
IR sources, the outflows are richly populated with water masers, which
indicates there is substantial ambient or infalling high-density material
broadly distributed within $\sim 100$ AU of the protostars.   Although the
extinction toward IRS\,3 is larger \citep{walther93}, the evolution of the
embedded protostar therein does not seem markedly less far along.

The detection of water maser emission tracing  compact disk components (10 to
20 AU long and $\sim 2$ AU wide), conical outflow, and wide-angle cavities is unusual.
These detections have required observations with the angular resolutions
characteristic of VLBI.  Disks traced by water masers on larger scales have
been inferred from   lower angular resolution data
\citep{shepherd99,torrelles97,torrelles98}, but confusion from  blending of
maser features is a risk in these cases, and confirmation requires higher
angular resolution observations.

The water maser features in IRS\,3 and IRS\,1 sample the underlying dense gas
incompletely, which limits our inferences. In order to solidify the
interpretation renewed long baseline interferometric observations are
necessary. Water maser emission in regions of low and intermediate mass star
formation is highly time variable
\citep[e.g.,][]{wilking94,torrelles98a}, and the accumulation of maps for many
epochs, registered and superposed, will better trace the underlying dense gas
structures around the protostars.  

\acknowledgements

We thank T. Beasley for his assistance in adjusting the data to a common
phase-center through reconstruction of the correlator model. We credit L.
Chernin for his initiation of the NGC\,2071IR proper motion study. This work
was supported in part by the Smithsonian Institution Scholarly Studies Program
and Smithsonian Astrophysical Observatory R\&D funds.

\newpage

\leftline{\bf Captions \hfill}

\noindent
Figure\,1--Comparison of total and imaged power for IRS\,1 and
IRS\,3 together.  {\it(bottom)} The total power spectrum for a roughly $\pm 1'$
field around NGC\,2071IR (solid line) and a channel-by-channel sum of the peak
flux density for each detected and fitted maser feature (dashed line) for 1996
March 24. {\it(top)} A ratio of the imaged and total power, which cannot
exceed unity.   
\bigskip

\noindent
Figure\,2--The distribution of water maser emission features in IRS\,3 at four
epochs in 1996 (see Table\,1) and their proper motions.  {\it (a)} All
features, where color indicates Doppler velocity.  Uncertainties in position
are typically $\ll 0\rlap{.}''001$.  Arrows indicate measured proper motions.
The lengths are scaled to show the expected changes in postion after 9 years. 
Uncertainties in proper motion magnitude and position angle ($1\sigma$) are
shown as tails on the arrows, where lengths indicate uncertainty in magnitude
and opening angle indicates uncertainty in position angle. The uncertainties
are too small to see in most cases.  The black cross and filled squares
indicate respectively the positions of the $\lambda1.3$\,cm radio continuum
peak and the water masers detected by Torrelles et al. (1998) with the VLA
$\sim 4$ months after the VLBA observations. The alignment of the VLA and VLBA
maps is uncertain by $0\rlap{.}''01$, as reflected in the size of the cross.
The map origin is $2\rlap{.}''32$ west and $5\rlap{.}''01$ north of the origin
for maps  of IRS\,1 (see Figure\,3). The red box indicates the region plotted
on an expanded scale below. {\it (b)} An enlargement that better shows the 
masers
associated with the putative protostellar disk and innermost portion of the
proposed outflow cone. The arrows correspond to the expected change in feature
postions after 1.5 years.

\bigskip

\noindent
Figure\,3--The distribution of water maser emission features in IRS\,1.
{\it(a)} All features, color coded to indicate Doppler velocity. 
The proper motion arrows
show the expected change in feature positions over 9 years.  The tails on the
proper motions here represent 3$\sigma$ uncertainties. The $\lambda1.3$\,cm
continuum peak (black cross) and maser features (colored squares) measured by
Torrelles et al. (1998) with the VLA are shown with  $0\rlap{.}''01$
uncertainty in alignment relative to the VLBA maps. The origin of the map is
$\alpha_{J2000}=5^h 47^m 04\rlap{.}^s7774$, $\delta_{J2000}=0^{\circ} 21'
42\rlap{.}''803$ with an uncertainty of $0\rlap{.}''05$ in both
coordinates.  The red boxes show
which fields are shown in the enlargements below.  {\it (b)} An enlargement
that better shows the masers associated with the putative protostellar
disk.  The arrows correspond to the expected change in feature
postions after 1.5 years. 
{\it (c)} The arc of maser features to the west that does not appear to
participate in the high-speed outflow, despite its apparent proximity. 

\bigskip

\noindent
Figure\,4--Components of our proposed models for {\it (a)} IRS\,3 and
{\it (b)} IRS\,1.

%\bibliography{ref_ngc2071}
%\bibliographystyle{apj}

\clearpage

\begin{deluxetable}{lclrccccc}
\tablewidth{6.4in}
\tablecaption{Journal of 1996 VLBA Observations}
\tablehead{
     \colhead{Date}  &
     \colhead{Track} &
     \colhead{Beam}  &
     \colhead{PA}    & 
     \colhead{RMS noise} & 
     \colhead{Maser Spots} &
     \colhead{Maser Features} \\
     \colhead{}      &
     \colhead{(h)}   &
     \colhead{(mas)} &
     \colhead{($^\circ$)} &
     \colhead{(Jy)}  &
     \colhead{}      &
     \colhead{} 
}

\startdata
March 24 & 8  & $0.92 \times 0.51$ & -9 & 0.014 & 337 & 74   \\   %112 ftrs w/o editing
June   2 & 8  & $1.08 \times 0.59$ & -1 & 0.016 & 259 & 67  \\   %88 ftrs w/o editing
July   5 & 8  & $1.03 \times 0.57$ &  8 & 0.016 & 310 & 81   \\   %116 ftrs w/o editing
August 3 & 8.5& $1.36 \times 1.26$ & 24 & 0.023 & 168 & 47   \\   %53 ftrs w/o editing
\enddata
\end{deluxetable}

\begin{deluxetable}{cccclrlrlrlcclc}
\tablewidth{0pt}
\rotate
\tabletypesize{\scriptsize}
\tablecaption{Water Maser Features in NGC2071\,IRS1 \& IRS3}
\tablehead{
\colhead{Object} & \colhead{Epoch} & \colhead{Feature} & 
\colhead{V$_{\rm LOS}$} & \colhead{Unc.} & \colhead{Peak} &
\colhead{Unc.} & \colhead{E-W} &
\colhead{Unc.} & \colhead{N-S} & 
\colhead{Unc.} & \colhead{Spots in} &
\colhead{HPFW\tablenotemark{a}} & \colhead{Unc.} &
\colhead{Motion} \\

\colhead{} & \colhead{No.} & \colhead{No.} & 
\colhead{} & \colhead{} & \colhead{F$_\nu$} &
\colhead{} & \colhead{Offset} &
\colhead{} & \colhead{Offset} & 
\colhead{} & \colhead{Feature} &
\colhead{} & \colhead{} &
\colhead{No.\tablenotemark{b}} \\

\colhead{} & \colhead{} & \colhead{} & 
\colhead{(\kms)} & \colhead{(\kms)} & \colhead{(Jy)} &
\colhead{(Jy)} & \colhead{(mas)} &
\colhead{(mas)} & \colhead{(mas)} & 
\colhead{(mas)} & \colhead{} &
\colhead{(\kms)} & \colhead{} &
\colhead{} \\
}
\startdata
IRS3 &  1  &   0   &  18.86 & 0.55   &  0.238 &   0.042 &   49.527    &  0.043 &     1.938   &    0.181 &    4   &    1.950 &   1.124 &   7  \\
IRS3 &  1  &   1   &  18.49 & 0.21   &  0.174 &   0.010 &   49.753    &  0.132 &     2.281   &    0.053 &    2   &          &         &      \\
IRS3 &  1  &   2   &  15.39 & 0.01   & 42.556 &   1.094 &   81.567    &  0.020 &    -0.891   &    0.066 &   18   &    0.881 &   0.024 &      \\
IRS3 &  1  &   3   &  15.22 & 0.21   &  0.320 &   0.018 &    7.851    &  0.042 &     9.846   &    0.029 &    3   &    0.857 &   0.040 &      \\
IRS3 &  1  &   4   &  13.17 & 0.21   &  0.095 &   0.014 &    2.609    &  0.176 &    41.282   &    0.173 &    1   &          & 	       &      \\
IRS3 &  1  &   5   &  12.75 & 0.21   &  0.122 &   0.014 &  104.294    &  0.036 &  -234.312   &    0.079 &    1   &          & 	       &      \\
IRS3 &  1  &   6   &  12.75 & 0.21   &  0.157 &   0.014 &  103.908    &  0.065 &  -236.052   &    0.099 &    1   &          & 	       &      \\
IRS3 &  1  &   7   &  11.43 & 0.05   &  1.437 &   0.093 &    2.741    &  0.269 &    40.831   &    0.084 &    7   &    1.520 &   0.113 &   3  \\
IRS3 &  1  &   8   &  10.64 & 0.21   &  0.154 &   0.014 &    1.485    &  0.049 &    40.768   &    0.045 &    1   &          & 	       &      \\
IRS3 &  1  &   9   &   7.87 & 0.03   &  1.583 &   0.084 &    0.196    &  0.114 &    43.502   &    0.142 &    7   &    1.264 &   0.077 &      \\
IRS3 &  1  &   10  &   6.70 & 0.09   &  0.575 &   0.035 &   -0.979    &  0.034 &    42.070   &    0.110 &    4   &    1.629 &   0.163 &      \\
IRS3 &  1  &   11  &   6.04 & 0.21   &  0.199 &   0.015 &    0.850    &  0.101 &     2.109   &    0.057 &    3   &    1.234 &   0.043 &      \\
IRS3 &  1  &   12  &   5.52 & 0.01   &  5.184 &   0.106 &   -5.224    &  0.169 &     1.056   &    0.082 &    5   &    0.776 &   0.019 &      \\
IRS3 &  1  &   13  &   5.24 & 0.21   &  1.601 &   0.010 &   -2.424    &  0.025 &     2.597   &    0.012 &    2   &          &         &      \\
IRS3 &  1  &   14  &   4.74 & 0.21   &  0.258 &   0.017 &   -6.361    &  0.015 &     7.660   &    0.027 &    1   &          & 	       &      \\
IRS3 &  1  &   15  &   4.44 & 0.01   & 60.097 &   1.094 &   -1.603    &  0.035 &     1.485   &    0.057 &    8   &    0.620 &   0.012 &      \\
IRS3 &  1  &   16  &   4.43 & 0.02   &  5.427 &   0.305 &   -3.969    &  0.070 &     5.032   &    0.070 &    4   &    0.617 &   0.041 &      \\
IRS3 &  1  &   17  &   3.98 & $<$0.01  & 44.358 &   0.529 &   -2.596    &  0.036 &     4.364   &    0.061 &   13   &    0.665 &   0.009 &      \\
IRS3 &  1  &   18  &   3.37 & 0.03   &  1.920 &   0.138 &   -7.509    &  0.213 &     7.701   &    0.063 &    4   &    0.808 &   0.073 &   6  \\
IRS3 &  1  &   19  &   3.34 & 0.21   &  0.279 &   0.010 &   -1.180    &  0.018 &     44.481  &    0.045 &    2   &          &         &      \\
IRS3 &  1  &   20  &   3.12 & 0.03   &  0.830 &   0.039 &    2.074    &  0.027 &     53.145  &    0.083 &    5   &    1.218 &   0.069 &      \\
IRS3 &  1  &   21  &   2.93 & 0.09   &  0.698 &   0.079 &   -7.983    &  0.043 &      8.895  &    0.084 &    4   &    1.217 &   0.164 &   7  \\
IRS3 &  1  &   22  &   2.62 & $<$0.01  &  0.669 &   0.007 &   -2.238    &  0.114 &     41.938  &    0.058 &    4   &    1.417 &   0.004 &      \\
IRS3 &  1  &   23  &   2.21 & 0.21   &  0.699 &   0.008 &    6.820    &  0.039 &     -2.838  &    0.164 &    3   &    0.808 &   0.005 &      \\
IRS3 &  1  &   24  &  -4.74 & 0.21   &  0.196 &   0.010 &  -41.687    &  0.040 &    204.994  &    0.060 &    2   &          &         &      \\
     &	   &	   &	    & 		    &	          & 	 & 	       &        & 	      & 	 & 	  & 	     & 	       &      \\
IRS3 &  2  &   0   &  25.25 & 0.21   &  0.583 &   0.073 &   -165.872  &  0.036 &   -95.223   &    0.016 &   3    &   3.317  &  0.023  &      \\
IRS3 &  2  &   1   &  24.97 & 0.21   &  0.244 &   0.016 &   -165.221  &  0.020 &   -95.783   &    0.035 &   1    &          &         &  0   \\
IRS3 &  2  &   2   &  16.72 & 0.01   &  0.314 &   0.008 &      8.780  &  0.058 &    12.176   &    0.424 &   4    &   1.238  &  0.011  &      \\
IRS3 &  2  &   3   &  16.07 & 0.21   &  0.119 &   0.009 &     52.978  &  0.082 &     0.134   &    0.086 &   3    &   0.936  &  0.044  &      \\
IRS3 &  2  &   4   &  15.84 & 0.21   &  0.116 &   0.011 &     83.502  &  0.077 &    -1.874   &    0.171 &   2    &          &         &  1   \\
IRS3 &  2  &   5   &  14.85 & $<$0.01  &  1.304 &   0.007 &     83.964  &  0.069 &    -2.008   &    0.070 &   6    &   0.894  &  0.003  &      \\
IRS3 &  2  &   6   &  13.59 & 0.21   &  0.206 &   0.016 &    107.363  &  0.024 &  -233.794   &    0.048 &   1    &          &         &  2   \\
IRS3 &  2  &   7   &  13.17 & 0.21   &  0.239 &   0.016 &    107.115  &  0.026 &  -234.631   &    0.050 &   1    &          & 	       &      \\
IRS3 &  2  &   8   &  12.71 & 0.21   &  0.139 &   0.015 &      5.726  &  0.085 &    40.833   &    0.134 &   3    &   1.294  &  0.062  &      \\
IRS3 &  2  &   9   &  11.87 & 0.01   &  1.095 &   0.017 &      4.858  &  0.071 &    40.648   &    0.079 &   8    &   1.511  &  0.026  &  8   \\
IRS3 &  2  &   10  &  11.87 & 0.02   &  0.160 &   0.008 &      7.905  &  0.107 &    39.217   &    0.070 &   4    &   1.497  &  0.021  &      \\
IRS3 &  2  &   11  &   5.61 & $<$0.01  &  1.770 &   0.007 &      8.498  &  0.069 &    43.646   &    0.012 &   6    &   1.142  &  0.002  &      \\
IRS3 &  2  &   12  &   5.58 & 0.21   &  0.113 &   0.016 &     -7.762  &  0.058 &     5.628   &    0.089 &   1    &          & 	       &      \\
IRS3 &  2  &   13  &   4.81 & 0.21   &  0.651 &   0.009 &      1.524  &  0.265 &     1.988   &    0.198 &   3    &   0.878  &  0.055  &  4   \\
IRS3 &  2  &   14  &   4.74 & 0.21   &  0.084 &   0.016 &     -6.095  &  0.069 &     8.231   &    0.104 &   1    &          & 	       &      \\
IRS3 &  2  &   15  &   4.32 & 0.21   &  0.157 &   0.016 &      4.748  &  0.036 &     0.124   &    0.055 &   1    &          & 	       &      \\
IRS3 &  2  &   16  &   4.17 & 0.21   &  0.624 &   0.018 &      2.986  &  0.087 &     1.403   &    0.012 &   3    &          &         &  5   \\
IRS3 &  2  &   17  &   2.71 & $<$0.01  &  1.003 &   0.007 &     -8.354  &  0.260 &     9.952   &    0.360 &   5    &   1.113  &  0.004  &      \\
IRS3 &  2  &   18  &   2.39 & 0.21   &  0.178 &   0.011 &      7.857  &  0.106 &    -3.364   &    0.093 &   2    &          &         &      \\
IRS3 &  2  &   19  &   0.00 & 0.02   & 12.255 &   0.686 &     -7.509  &  0.072 &     7.701   &    0.042 &   4    &   0.607  &  0.039  &  6   \\
IRS3 &  2  &   20  &  -0.46 & 0.21   &  0.296 &   0.011 &     -3.844  &  0.015 &    58.999   &    0.124 &   2    &          &         &      \\
IRS3 &  2  &   21  &  -3.14 & 0.01   &  0.525 &   0.008 &    -42.436  &  0.046 &   233.593   &    0.173 &   4    &   0.968  &  0.007  &      \\
IRS3 &  2  &   22  &  -7.71 & 0.21   &  0.106 &   0.011 &    -39.831  &  0.037 &   243.294   &    0.117 &   2    &          &         &      \\
IRS3 &  2  &   23  &  -7.90 & 0.21   &  0.099 &   0.016 &     15.138  &  0.048 &    43.676   &    0.087 &   1    &          & 	       &      \\
     &	   &	   &	    & 		    &	          & 	 & 	       &        & 	      & 	 & 	  & 	     & 	       &      \\
IRS3 &  3  &   0   &  24.82 & 0.21   & 0.176  &  0.011  &   -166.276  &  0.042 &   -97.725   &   0.048  &   2    &          &         &  0   \\
IRS3 &  3  &   1   &  24.55 & 0.21   & 0.149  &  0.016  &   -166.491  &  0.031 &   -97.364   &   0.055  &   1    &          & 	       &      \\
IRS3 &  3  &   2   &  21.74 & 0.21   & 0.097  &  0.011  &   -364.672  &  0.100 &   -72.643   &   0.123  &   2    &          &         &      \\
IRS3 &  3  &   3   &  16.69 & 0.02   & 2.661  &  0.164  &      8.615  &  0.048 &    13.660   &   0.083  &   5    &   0.715  &  0.044  &      \\
IRS3 &  3  &   4   &  16.54 & 0.21   & 0.122  &  0.017  &      5.759  &  0.057 &    11.830   &   0.073  &   1    &          & 	       &      \\
IRS3 &  3  &   5   &  16.33 & 0.01   & 2.784  &  0.056  &      8.971  &  0.096 &    12.876   &   0.230  &   5    &   0.988  &  0.022  &      \\
IRS3 &  3  &   6   &  16.22 & 0.21   & 0.174  &  0.018  &      7.773  &  0.073 &    10.812   &   0.256  &   3    &          &         &      \\
IRS3 &  3  &   7   &  15.28 & 0.21   & 0.088  &  0.016  &     84.457  &  0.077 &    -2.549   &   0.128  &   1    &          &         &  1   \\
IRS3 &  3  &   8   &  15.08 & 0.21   & 0.110  &  0.011  &      9.445  &  0.034 &    11.291   &   0.110  &   2    &          &         &      \\
IRS3 &  3  &   9   &  14.86 & 0.21   & 0.163  &  0.009  &     85.027  &  0.047 &    -2.624   &   0.097  &   3    &   0.880  &  0.021  &      \\
IRS3 &  3  &   10  &  13.59 & 0.21   & 0.113  &  0.016  &      5.189  &  0.094 &    40.256   &   0.075  &   1    &          & 	       &      \\
IRS3 &  3  &   11  &  13.37 & 0.21   & 0.172  &  0.011  &    107.709  &  0.052 &  -236.887   &   0.042  &   2    &          &         &      \\
IRS3 &  3  &   12  &  13.32 & 0.21   & 0.254  &  0.011  &    108.169  &  0.051 &  -234.600   &   0.061  &   2    &          &         &  2   \\
IRS3 &  3  &   13  &  11.61 & 0.01   & 0.580  &  0.007  &      7.078  &  0.265 &    40.742   &   0.044  &   5    &   1.354  &  0.022  &  3   \\
IRS3 &  3  &   14  &  11.16 & 0.21   & 0.316  &  0.058  &      7.247  &  0.034 &    39.761   &   0.070  &   3    &   1.021  &  0.121  &      \\
IRS3 &  3  &   15  &   8.27 & 0.21   & 0.135  &  0.011  &     72.954  &  0.068 &    22.755   &   0.083  &   2    &          &         &      \\
IRS3 &  3  &   16  &   8.11 & 0.21   & 0.135  &  0.024  &     72.971  &  0.081 &    22.782   &   0.107  &   1    &          & 	       &      \\
IRS3 &  3  &   17  &   7.43 & 0.04   & 3.028  &  0.138  &      5.934  &  0.051 &    40.842   &   0.093  &  12    &   2.135  &  0.114  &      \\
IRS3 &  3  &   18  &   6.03 & 0.21   & 0.211  &  0.009  &      3.857  &  0.131 &     1.298   &   0.083  &   3    &   0.784  &  0.016  &      \\
IRS3 &  3  &   19  &   5.85 & 0.21   & 0.137  &  0.011  &     -4.296  &  0.086 &     1.311   &   0.055  &   2    &          &         &      \\
IRS3 &  3  &   20  &   5.58 & 0.21   & 0.098  &  0.016  &      0.516  &  0.148 &     2.294   &   0.107  &   1    &          & 	       &      \\
IRS3 &  3  &   21  &   4.76 & 0.21   & 0.427  &  0.010  &      1.095  &  0.027 &     2.129   &   0.063  &   3    &   1.009  &  0.023  &  4   \\
IRS3 &  3  &   22  &   4.63 & 0.21   & 0.610  &  0.009  &      3.850  &  0.022 &     0.798   &   0.036  &   3    &   0.738  &  0.055  &      \\
IRS3 &  3  &   23  &   4.62 & 0.21   & 1.224  &  0.081  &     -0.934  &  0.024 &     0.742   &   0.105  &   3    &   0.691  &  0.048  &      \\
IRS3 &  3  &   24  &   4.40 & 0.02   & 3.849  &  0.149  &     -0.011  &  0.127 &    -0.046   &   0.039  &   6    &   0.814  &  0.040  &      \\
IRS3 &  3  &   25  &   4.33 & $<$0.01  & 3.547  &  0.007  &      2.991  &  0.137 &     1.309   &   0.178  &   6    &   0.691  &  0.002  &  5   \\
IRS3 &  3  &   26  &   4.32 & 0.21   & 0.109  &  0.016  &      4.967  &  0.074 &    -0.168   &   0.084  &   1    &          & 	       &      \\
IRS3 &  3  &   27  &   4.09 & 0.21   & 0.163  &  0.011  &     -1.453  &  0.040 &     4.064   &   0.119  &   2    &          &         &      \\
IRS3 &  3  &   28  &   2.90 & 0.03   & 6.170  &  0.684  &     -7.509  &  0.098 &     7.701   &   0.131  &   6    &   0.654  &  0.071  &  6   \\
IRS3 &  3  &   29  &   2.85 & 0.21   & 0.164  &  0.011  &    -10.575  &  0.055 &     6.340   &   0.058  &   2    &          &         &      \\
IRS3 &  3  &   30  &   2.65 & 0.21   & 0.569  &  0.009  &     -8.555  &  0.122 &     9.940   &   0.196  &   3    &   0.858  &  0.010  &  7   \\
IRS3 &  3  &   31  &   2.21 & 0.21   & 0.127  &  0.019  &      2.379  &  0.090 &    -1.042   &   0.090  &   1    &          & 	       &      \\
IRS3 &  3  &   32  &   2.21 & 0.21   & 0.108  &  0.019  &     -4.586  &  0.091 &     5.929   &   0.120  &   1    &          & 	       &      \\
IRS3 &  3  &   33  &  -1.47 & 3.73   & 0.087  &  0.019  &     -4.078  &  0.079 &    60.835   &   0.193  &   4    &   3.810  &  5.656  &      \\
IRS3 &  3  &   34  &  -3.32 & 0.21   & 0.240  &  0.034  &    -42.563  &  0.089 &   233.766   &   0.378  &   3    &   0.511  &  0.244  &      \\
     &	   &	   &	    & 		    &	          & 	 & 	       &        & 	      & 	 & 	  & 	     & 	       &      \\
IRS3 &  4  &   0   &  27.07 & 0.21   & 0.151  &  0.023  &   -576.317  &   0.096 &    351.813 &    0.104 &    1	  & 	     & 	       &      \\
IRS3 &  4  &   1   &  24.78 & 0.21   & 0.234  &  0.016  &   -167.292  &   0.048 &    -98.884 &    0.062 &    2   &          &         &  0   \\
IRS3 &  4  &   2   &  23.08 & 0.21   & 0.133  &  0.016  &   -174.438  &   0.082 &   -103.376 &    0.088 &    2   &          &         &      \\
IRS3 &  4  &   3   &  15.86 & $<$0.01  & 1.386  &  0.012  &      9.209  &   0.043 &     12.410 &    0.058 &    4   &   0.825  &  0.003  &      \\	
IRS3 &  4  &   4   &  13.17 & 0.21   & 0.197  &  0.023  &    108.934  &   0.074 &   -235.332 &    0.099 &    1   &          &         &  2   \\
IRS3 &  4  &   5   &   7.49 & 0.21   & 0.157  &  0.016  &     17.116  &   0.068 &    -22.903 &    0.148 &    2   &          &         &      \\	
IRS3 &  4  &   6   &   5.43 & 0.21   & 0.138  &  0.016  &      4.182  &   0.164 &     37.072 &    0.101 &    2   &          &         &      \\	
IRS3 &  4  &   7   &   5.39 & 0.21   & 0.100  &  0.016  &      4.640  &   0.135 &     39.025 &    0.192 &    2   &          &         &      \\	
IRS3 &  4  &   8   &   4.91 & 0.21   & 0.242  &  0.016  &      0.757  &   0.051 &      2.215 &    0.063 &    2   &          &         &  4   \\
IRS3 &  4  &   9   &   4.64 & 0.21   & 0.718  &  0.013  &     -0.282  &   0.075 &      0.237 &    0.026 &    3   &   0.782  &  0.065  &      \\	
IRS3 &  4  &   10  &   4.60 & 0.21   & 0.165  &  0.016  &     -1.873  &   0.135 &      6.901 &    0.091 &    2   &          &         &      \\	
IRS3 &  4  &   11  &   4.41 & 0.03   & 1.177  &  0.090  &      3.102  &   0.098 &      1.218 &    0.110 &    4   &   0.723  &  0.066  &  5   \\
IRS3 &  4  &   12  &   3.92 & 0.03   & 2.996  &  0.157  &      3.119  &   0.172 &     54.334 &    0.384 &   10   &   1.242  &  0.076  &      \\	
IRS3 &  4  &   13  &   3.17 & 0.02   & 2.347  &  0.121  &     -7.509  &   0.225 &      7.701 &    0.197 &    6   &   0.942  &  0.058  &  6   \\
IRS3 &  4  &   14  &   2.61 & 0.21   & 0.518  &  0.013  &     -8.886  &   0.124 &     10.257 &    0.292 &    3   &   5.375  &  0.009  &  7   \\
     &	   &	   &	    & 		    &	          & 	 & 	       &        & 	      & 	 & 	  & 	     & 	       &      \\
IRS1 &  1  &   0   &  27.31 & 0.02   & 1.356  &  0.050  &   306.516   &  0.085 &     67.226  &    0.106 &   6    &   0.928  &  0.039  &      \\
IRS1 &  1  &   1   &  26.65 & 0.21   & 0.291  &  0.014  &   313.172   &  0.034 &     65.171  &    0.048 &   1    &          & 	       &      \\
IRS1 &  1  &   2   &  26.44 & 0.21   & 0.191  &  0.010  &    -8.852   &  0.013 &    -37.273  &    0.023 &   2    &          &         &      \\
IRS1 &  1  &   3   &  26.23 & 0.21   & 0.220  &  0.014  &   311.196   &  0.156 &     66.112  &    0.107 &   1    &          & 	       &      \\
IRS1 &  1  &   4   &  18.22 & 0.21   & 0.100  &  0.014  &     4.440   &  0.086 &    -23.098  &    0.087 &   1    &          & 	       &      \\
IRS1 &  1  &   5   &  18.02 & 0.03   & 4.670  &  0.326  &   251.866   &  0.120 &   -269.003  &    0.282 &   6    &   0.877  &  0.071  &      \\
IRS1 &  1  &   6   &  17.80 & 0.21   & 0.106  &  0.014  &     4.432   &  0.066 &    -23.593  &    0.132 &   1    &          & 	       &      \\
IRS1 &  1  &   7   &  17.16 & 0.21   & 0.250  &  0.010  &   -58.338   &  0.231 &    -76.076  &    0.092 &   2    &          &         &      \\
IRS1 &  1  &   8   &  17.10 & 0.08   & 0.509  &  0.060  &     5.063   &  0.172 &    -23.950  &    0.158 &   4    &   0.964  &  0.178  &      \\
IRS1 &  1  &   9   &  16.98 & 0.21   & 0.270  &  0.014  &   -55.163   &  0.122 &    -74.003  &    0.135 &   3    &          &         &      \\
IRS1 &  1  &   10  &  16.96 & 0.21   & 0.110  &  0.014  &     6.371   &  0.091 &      1.210  &    0.204 &   1    &          & 	       &      \\
IRS1 &  1  &   11  &  16.73 & 0.21   & 0.135  &  0.010  &   236.246   &  0.258 &     73.271  &    0.082 &   2    &          &         &      \\
IRS1 &  1  &   12  &  16.54 & 0.21   & 0.122  &  0.014  &   236.999   &  0.066 &     73.570  &    0.067 &   1    &          & 	       &      \\
IRS1 &  1  &   13  &  16.54 & 0.21   & 0.067  &  0.014  &   238.075   &  0.072 &     73.798  &    0.111 &   1    &          & 	       &      \\
IRS1 &  1  &   14  &  15.70 & 0.21   & 0.080  &  0.015  &   125.021   &  0.042 &   -145.912  &    0.077 &   1    &          & 	       &      \\
IRS1 &  1  &   15  &  15.28 & 0.21   & 0.097  &  0.015  &     3.551   &  0.038 &    -18.009  &    0.063 &   1    &          & 	       &      \\
IRS1 &  1  &   16  &  15.28 & 0.21   & 0.037  &  0.015  &     4.020   &  0.091 &     -1.314  &    0.165 &   1    &          & 	       &      \\
IRS1 &  1  &   17  &  14.78 & 0.21   & 0.321  &  0.039  &     3.346   &  0.101 &     -1.673  &    0.084 &   3    &   2.961  &  0.073  &      \\
IRS1 &  1  &   18  &  14.26 & 0.21   & 0.880  &  0.014  &   240.638   &  0.053 &     64.319  &    0.008 &   3    &          &         &      \\
IRS1 &  1  &   19  &  14.01 & 0.21   & 0.326  &  0.008  &     3.004   &  0.018 &     -2.110  &    0.068 &   3    &   0.075  &  0.049  &      \\
IRS1 &  1  &   20  &  13.56 & 0.01   & 2.007  &  0.067  &   245.521   &  0.021 &     75.425  &    0.101 &   4    &   0.594  &  0.022  &   3  \\
IRS1 &  1  &   21  &  12.52 & 0.03   & 0.549  &  0.033  &   -86.590   &  0.082 &    -99.246  &    0.086 &   4    &   1.018  &  0.078  &   5  \\
IRS1 &  1  &   22  &  12.49 & 0.21   & 0.236  &  0.010  &     2.340   &  0.024 &      1.539  &    0.045 &   2    &          &         &      \\
IRS1 &  1  &   23  &  12.44 & 0.01   & 0.899  &  0.007  &    80.769   &  0.017 &   -112.261  &    0.245 &   4    &   1.225  &  0.012  &      \\
IRS1 &  1  &   24  &  12.07 & 0.05   & 1.119  &  0.100  &     2.847   &  0.064 &      3.671  &    0.015 &   5    &   1.029  &  0.111  &   8  \\
IRS1 &  1  &   25  &  11.90 & 0.21   & 0.274  &  0.014  &    82.365   &  0.095 &   -112.212  &    0.084 &   1    &          &         &   7  \\
IRS1 &  1  &   26  &  11.22 & 0.02   & 3.062  &  0.165  &    -1.078   &  0.081 &     -3.972  &    0.060 &   6    &   0.734  &  0.049  &      \\
IRS1 &  1  &   27  &  10.96 & 0.21   & 0.360  &  0.014  &    -0.132   &  0.024 &     -0.209  &    0.095 &   3    &          &         &      \\
IRS1 &  1  &   28  &  10.55 & 0.21   & 0.292  &  0.015  &  -238.946   &  0.121 &    -51.450  &    0.135 &   3    &          &         &  10  \\
IRS1 &  1  &   29  &  10.29 & $<$0.01  & 1.098  &  0.007  &  -527.275   &  0.103 &      1.361  &    0.323 &   4    &   0.847  &  0.004  &      \\
IRS1 &  1  &   30  &   9.80 & 0.21   & 0.295  &  0.014  &  -338.989   &  0.092 &    -52.062  &    0.072 &   1    &          & 	       &      \\
IRS1 &  1  &   31  &   9.11 & $<$0.01  & 1.722  &  0.019  &  -280.679   &  0.057 &    -96.833  &    0.238 &   4    &   0.807  &  0.011  &      \\
IRS1 &  1  &   32  &   8.96 & 0.21   & 0.134  &  0.014  &  -279.229   &  0.032 &    -88.479  &    0.054 &   1    &          &         &  14  \\
IRS1 &  1  &   33  &   8.96 & 0.21   & 0.122  &  0.014  &  -526.109   &  0.028 &      5.261  &    0.050 &   1    &          & 	       &      \\
IRS1 &  1  &   34  &   8.85 & 0.21   & 1.743  &  0.121  &  -280.828   &  0.055 &    -92.521  &    0.060 &   3    &   0.762  &  0.067  &      \\
IRS1 &  1  &   35  &   8.53 & 0.21   & 0.109  &  0.014  &  -331.589   &  0.031 &    -55.083  &    0.056 &   1    &          &         &  15  \\
IRS1 &  1  &   36  &   8.53 & 0.21   & 0.077  &  0.014  &  -526.789   &  0.111 &      5.021  &    0.089 &   1    &          & 	       &      \\
IRS1 &  1  &   37  &   7.99 & 0.21   & 0.980  &  0.014  &  -273.911   &  0.064 &   -107.442  &    0.070 &   3    &          &         &      \\
IRS1 &  1  &   38  &   7.37 & 0.09   & 0.283  &  0.024  &    -0.601   &  0.254 &     13.623  &    0.228 &   4    &   1.422  &  0.212  &      \\
IRS1 &  1  &   39  &   7.16 & $<$0.01  & 0.600  &  0.007  &    15.099   &  0.148 &     34.564  &    0.098 &   4    &   1.381  &  0.005  &      \\
IRS1 &  1  &   40  &   6.43 & 0.21   & 0.235  &  0.014  &    -1.685   &  0.115 &     15.246  &    0.117 &   1    &          & 	       &      \\
IRS1 &  1  &   41  &   5.98 & 0.21   & 0.301  &  0.018  &    -0.978   &  0.039 &     17.810  &    0.197 &   3    &   1.077  &  0.109  &      \\
IRS1 &  1  &   42  &   4.94 & 0.21   & 0.363  &  0.010  &    52.420   &  0.134 &     86.622  &    0.076 &   2    &          &         &  19  \\
IRS1 &  1  &   43  &   4.74 & 0.21   & 0.205  &  0.015  &   112.329   &  0.093 &    -30.362  &    0.096 &   1    &          & 	       &      \\
IRS1 &  1  &   44  &   3.97 & 0.21   & 0.714  &  0.015  &   111.716   &  0.109 &    -31.054  &    0.036 &   3    &          &         &      \\
IRS1 &  1  &   45  &   3.81 & 0.03   & 2.147  &  0.157  &   113.572   &  0.085 &    -33.072  &    0.036 &   4    &   0.813  &  0.064  &      \\
IRS1 &  1  &   46  &   3.74 & 0.21   & 1.278  &  0.010  &   113.093   &  0.071 &    -32.872  &    0.055 &   2    &          &         &      \\
IRS1 &  1  &   47  &   0.86 & 0.07   & 0.628  &  0.038  &  -389.197   &  0.031 &    -86.407  &    0.027 &   5    &   1.643  &  0.156  &      \\
IRS1 &  1  &   48  &  -1.33 & 0.01   &25.678  &  0.618  &    18.767   &  0.078 &     46.433  &    0.031 &  11    &   0.759  &  0.021  &  21  \\
     &	   &	   &	    & 		    &	          & 	 & 	       &        & 	      & 	 & 	  & 	     & 	       &      \\
IRS1 &  2  &   0   &  27.48 & 0.21   & 0.346  &  0.009  &   312.352   &  0.065 &     67.075  &    0.199 &   3    &   0.029  &  0.019  &      \\
IRS1 &  2  &   1   &  26.90 & 0.01   & 2.625  &  0.095  &   310.912   &  0.106 &     68.016  &    0.107 &   5    &   0.714  &  0.031  &   0  \\
IRS1 &  2  &   2   &  26.85 & 0.21   & 0.094  &  0.011  &   313.495   &  0.068 &     67.057  &    0.107 &   2    &          &         &      \\
IRS1 &  2  &   3   &  26.23 & 0.21   & 0.072  &  0.016  &   312.199   &  0.084 &     68.232  &    0.124 &   1    &          & 	       &      \\
IRS1 &  2  &   4   &  18.18 & 0.21   & 0.454  &  0.010  &     4.415   &  0.038 &    -10.777  &    0.034 &   3    &   0.694  &  0.017  &   1  \\
IRS1 &  2  &   5   &  18.03 & 0.21   & 0.235  &  0.011  &   251.438   &  0.154 &   -270.058  &    0.061 &   2    &          &         &      \\
IRS1 &  2  &   6   &  17.91 & 0.05   & 0.419  &  0.063  &   251.627   &  0.141 &   -269.824  &    0.114 &   4    &   0.676  &  0.116  &      \\
IRS1 &  2  &   7   &  17.63 & 0.02   & 0.185  &  0.008  &     6.350   &  0.092 &    -22.991  &    0.035 &   4    &   1.213  &  0.018  &      \\
IRS1 &  2  &   8   &  16.96 & 0.21   & 0.102  &  0.016  &   242.141   &  0.046 &     73.555  &    0.085 &   1    &          & 	       &      \\
IRS1 &  2  &   9   &  15.87 & 0.21   & 0.170  &  0.011  &   -82.765   &  0.028 &   -107.399  &    0.124 &   2    &          &         &      \\
IRS1 &  2  &   10  &  15.81 & 0.02   & 2.047  &  0.087  &   241.923   &  0.131 &     74.004  &    0.089 &   6    &   0.867  &  0.043  &      \\
IRS1 &  2  &   11  &  14.46 & $<$0.01  & 2.228  &  0.018  &   245.670   &  0.032 &     76.712  &    0.136 &   5    &   0.729  &  0.006  &      \\
IRS1 &  2  &   12  &  14.21 & 0.21   & 0.696  &  0.011  &   246.101   &  0.346 &     76.347  &    0.028 &   2    &          &         &   2  \\
IRS1 &  2  &   13  &  13.88 & 0.02   & 1.553  &  0.064  &   247.652   &  0.210 &     75.475  &    0.120 &   5    &   0.817  &  0.040  &   3  \\
IRS1 &  2  &   14  &  13.36 & 0.21   & 0.110  &  0.011  &   245.706   &  0.222 &     75.950  &    0.063 &   2    &          &         &      \\
IRS1 &  2  &   15  &  12.97 & 0.21   & 0.138  &  0.011  &  -113.776   &  0.035 &   -122.666  &    0.049 &   2    &          &         &   4  \\
IRS1 &  2  &   16  &  12.75 & 0.21   & 0.122  &  0.016  &   -85.875   &  0.039 &    -99.556  &    0.071 &   1    &          & 	       &      \\
IRS1 &  2  &   17  &  12.54 & 0.03   & 0.337  &  0.014  &   -86.617   &  0.116 &   -100.016  &    0.034 &   4    &   1.130  &  0.064  &   5  \\
IRS1 &  2  &   18  &  12.32 & 0.21   & 0.269  &  0.009  &    82.873   &  0.103 &   -113.788  &    0.044 &   3    &   0.149  &  0.080  &   6  \\
IRS1 &  2  &   19  &  12.03 & 0.02   & 6.692  &  0.176  &     1.928   &  0.104 &      2.232  &    0.115 &   9    &   1.240  &  0.035  &   8  \\
IRS1 &  2  &   20  &  11.15 & 0.21   & 0.429  &  0.016  &     0.754   &  0.030 &      0.865  &    0.209 &   3    &          &         &      \\
IRS1 &  2  &   21  &  10.67 & 0.21   & 0.184  &  0.009  &  -528.083   &  0.095 &      1.760  &    0.239 &   3    &   0.725  &  0.018  &      \\
IRS1 &  2  &   22  &  10.44 & 0.21   & 0.258  &  0.011  &    -0.886   &  0.183 &     -0.472  &    0.078 &   2    &          &         &   9  \\
IRS1 &  2  &   23  &   9.77 & 0.03   & 2.645  &  0.133  &    -1.715   &  0.025 &     -0.177  &    0.103 &   7    &   1.174  &  0.068  &      \\
IRS1 &  2  &   24  &   9.20 & 0.21   & 0.516  &  0.011  &  -281.157   &  0.018 &    -96.255  &    0.087 &   2    &          &         &  11  \\
IRS1 &  2  &   25  &   9.12 & 0.21   & 0.128  &  0.011  &  -338.458   &  0.130 &    -56.191  &    0.057 &   2    &          &         &      \\
IRS1 &  2  &   26  &   8.73 & 0.21   & 0.674  &  0.011  &  -278.492   &  0.082 &   -105.129  &    0.028 &   2    &          &         &  13  \\
IRS1 &  2  &   27  &   8.69 & 0.03   & 0.919  &  0.075  &  -333.954   &  0.102 &    -55.202  &    0.014 &   4    &   0.868  &  0.080  &  15  \\
IRS1 &  2  &   28  &   8.65 & $<$0.01  & 1.716  &  0.008  &  -280.958   &  0.185 &    -91.905  &    0.059 &   4    &   0.757  &  0.002  &      \\
IRS1 &  2  &   29  &   8.61 & 0.21   & 0.955  &  0.015  &  -279.556   &  0.137 &    -88.440  &    0.146 &   3    &   0.655  &  0.011  &  14  \\
IRS1 &  2  &   30  &   8.11 & 0.21   & 0.159  &  0.016  &  -277.961   &  0.030 &   -103.425  &    0.068 &   1    &          & 	       &      \\
IRS1 &  2  &   31  &   8.10 & 0.21   & 0.421  &  0.009  &  -339.638   &  0.088 &    -56.331  &    0.020 &   3    &   0.909  &  0.008  &  16  \\
IRS1 &  2  &   32  &   7.79 & 0.21   & 0.734  &  0.016  &  -277.603   &  0.375 &   -100.144  &    0.064 &   3    &          &         &      \\
IRS1 &  2  &   33  &   7.69 & 0.21   & 0.375  &  0.009  &  -274.551   &  0.269 &   -106.352  &    0.195 &   3    &   0.977  &  0.009  &  17  \\
IRS1 &  2  &   34  &   7.05 & 0.01   & 0.510  &  0.008  &    89.858   &  0.093 &    -34.312  &    0.021 &   4    &   0.822  &  0.015  &      \\
IRS1 &  2  &   35  &   6.46 & 0.21   & 0.312  &  0.029  &  -144.978   &  0.390 &   -135.359  &    0.056 &   3    &   1.221  &  0.243  &  18  \\
IRS1 &  2  &   36  &   6.40 & 0.03   & 0.407  &  0.021  &    14.407   &  0.140 &     34.923  &    0.145 &   5    &   1.214  &  0.077  &      \\
IRS1 &  2  &   37  &   4.98 & 0.21   & 0.235  &  0.011  &    52.508   &  0.022 &     86.286  &    0.126 &   2    &          &         &  19  \\
IRS1 &  2  &   38  &   3.97 & $<$0.01  & 6.190  &  0.007  &  -392.692   &  0.122 &    -87.572  &    0.066 &   6    &   0.808  &  0.001  &      \\
IRS1 &  2  &   39  &   3.74 & $<$0.01  & 3.397  &  0.009  &  -393.465   &  0.223 &    -86.993  &    0.148 &   6    &   0.909  &  0.003  &      \\
IRS1 &  2  &   40  &   3.68 & 0.21   & 0.259  &  0.011  &   114.580   &  0.101 &    -32.622  &    0.025 &   2    &          &         &      \\
IRS1 &  2  &   41  &   2.21 & 0.21   & 0.736  &  0.016  &  -408.801   &  0.013 &    -80.038  &    0.012 &   1    &          &         &  20  \\
IRS1 &  2  &   42  &  -1.22 & 0.03   & 3.251  &  0.221  &    19.799   &  0.131 &     47.060  &    0.207 &   8    &   0.896  &  0.067  &  21  \\
     &	   &	   &	    & 		    &	          & 	 & 	       &        & 	      & 	 & 	  & 	     & 	       &      \\
IRS1 &  3  &   0   &  26.95 & 0.12   & 0.383  &  0.138  &   311.282   &  0.161 &     68.804  &    0.177 &   4    &   0.637  &  0.290  &      \\
IRS1 &  3  &   1   &  26.83 & 0.04   & 1.038  &  0.087  &   311.514   &  0.029 &     68.880  &    0.112 &   4    &   0.728  &  0.085  &   0  \\
IRS1 &  3  &   2   &  18.49 & $<$0.01  & 1.760  &  0.007  &     4.297   &  0.074 &    -11.144  &    0.121 &   6    &   0.620  &  0.002  &   1  \\
IRS1 &  3  &   3   &  18.45 & 0.21   & 0.172  &  0.011  &     3.799   &  0.028 &    -11.403  &    0.039 &   2    &          &         &      \\
IRS1 &  3  &   4   &  18.22 & 0.21   & 0.134  &  0.017  &     2.211   &  0.034 &     -5.776  &    0.065 &   1    &          & 	       &      \\
IRS1 &  3  &   5   &  17.77 & 0.21   & 0.153  &  0.009  &     6.066   &  0.256 &    -23.114  &    0.077 &   3    &   0.878  &  0.044  &      \\
IRS1 &  3  &   6   &  16.01 & 0.21   & 0.373  &  0.045  &   242.997   &  0.015 &     74.057  &    0.041 &   3    &   0.830  &  0.139  &      \\
IRS1 &  3  &   7   &  15.78 & 0.21   & 0.202  &  0.028  &   -82.575   &  0.090 &   -107.785  &    0.057 &   3    &   1.707  &  0.113  &      \\
IRS1 &  3  &   8   &  14.33 & 0.21   & 0.949  &  0.009  &   247.245   &  0.213 &     76.327  &    0.119 &   3    &   2.730  &  0.004  &   2  \\
IRS1 &  3  &   9   &  14.07 & 0.09   & 0.432  &  0.079  &   248.804   &  0.383 &     75.383  &    0.089 &   4    &   0.925  &  0.205  &   3  \\
IRS1 &  3  &   10  &  13.53 & 0.21   & 0.089  &  0.015  &     0.083   &  0.083 &     -3.504  &    0.202 &   3    &          &         &      \\
IRS1 &  3  &   11  &  13.09 & 0.21   & 0.416  &  0.043  &  -113.611   &  0.092 &   -123.164  &    0.217 &   3    &   2.382  &  0.046  &   4  \\
IRS1 &  3  &   12  &  13.07 & 0.21   & 0.183  &  0.033  &     2.148   &  0.106 &      0.779  &    0.083 &   3    &   2.050  &  0.129  &      \\
IRS1 &  3  &   13  &  12.90 & 0.21   & 0.346  &  0.011  &   -85.993   &  0.156 &   -100.359  &    0.079 &   2    &          &         &      \\
IRS1 &  3  &   14  &  12.89 & 0.21   & 0.206  &  0.011  &  -109.787   &  0.041 &   -122.274  &    0.119 &   2    &          &         &      \\
IRS1 &  3  &   15  &  12.49 & 0.01   & 0.244  &  0.008  &    81.978   &  0.270 &   -113.749  &    0.161 &   4    &   1.147  &  0.014  &   6  \\
IRS1 &  3  &   16  &  12.15 & 0.21   & 0.177  &  0.011  &    80.835   &  0.038 &   -113.043  &    0.045 &   2    &          &         &   7  \\
IRS1 &  3  &   17  &  12.08 & $<$0.01  & 1.378  &  0.008  &     1.417   &  0.152 &      0.404  &    0.095 &   4    &   0.926  &  0.002  &   8  \\
IRS1 &  3  &   18  &  10.40 & 0.03   & 0.600  &  0.027  &    -0.942   &  0.145 &     -0.378  &    0.097 &   5    &   1.224  &  0.066  &   9  \\
IRS1 &  3  &   19  &  10.22 & 0.21   & 0.086  &  0.016  &  -281.251   &  0.090 &    -96.156  &    0.096 &   1    &          & 	       &      \\
IRS1 &  3  &   20  &  10.05 & 0.21   & 0.114  &  0.011  &  -241.839   &  0.040 &    -51.445  &    0.067 &   2    &          &         &  10  \\
IRS1 &  3  &   21  &   9.80 & 0.21   & 0.109  &  0.016  &  -279.737   &  0.049 &    -88.337  &    0.079 &   1    &          & 	       &      \\
IRS1 &  3  &   22  &   9.27 & 0.21   & 0.804  &  0.011  &  -281.153   &  0.041 &    -96.114  &    0.037 &   2    &          &         &  11  \\
IRS1 &  3  &   23  &   8.77 & 0.21   & 0.149  &  0.011  &  -335.154   &  0.049 &    -55.674  &    0.098 &   2    &          &         &  15  \\
IRS1 &  3  &   24  &   8.73 & 0.21   & 0.289  &  0.011  &  -278.484   &  0.023 &   -105.075  &    0.144 &   2    &          &         &  13  \\
IRS1 &  3  &   25  &   8.71 & 0.02   & 1.641  &  0.086  &  -281.020   &  0.232 &    -91.615  &    0.372 &   4    &   0.834  &  0.056  &      \\
IRS1 &  3  &   26  &   8.62 & 0.21   & 1.190  &  0.087  &  -279.674   &  0.143 &    -88.512  &    0.093 &   3    &   0.806  &  0.079  &  15  \\
IRS1 &  3  &   27  &   8.53 & 0.21   & 0.147  &  0.023  &  -342.177   &  0.035 &    -50.918  &    0.080 &   1    &          & 	       &      \\
IRS1 &  3  &   28  &   8.53 & 0.21   & 0.131  &  0.023  &  -343.258   &  0.039 &    -57.964  &    0.071 &   1    &          & 	       &      \\
IRS1 &  3  &   29  &   8.32 & 0.21   & 0.206  &  0.011  &  -277.948   &  0.168 &   -103.726  &    0.055 &   2    &          &         &      \\
IRS1 &  3  &   30  &   8.17 & $<$0.01  & 8.162  &  0.006  &  -340.117   &  0.103 &    -56.404  &    0.071 &   9    &   0.882  &  0.001  &  16  \\
IRS1 &  3  &   31  &   7.69 & 0.21   & 0.175  &  0.019  &  -342.198   &  0.026 &    -50.873  &    0.056 &   1    &          & 	       &      \\
IRS1 &  3  &   32  &   7.50 & 0.21   & 0.480  &  0.011  &  -274.422   &  0.110 &   -106.039  &    0.014 &   2    &          &         &  17  \\
IRS1 &  3  &   33  &   6.62 & 0.21   & 0.212  &  0.011  &  -144.569   &  0.314 &   -135.441  &    0.031 &   2    &          &         &  18  \\
IRS1 &  3  &   34  &   5.35 & 0.21   & 0.077  &  0.011  &  -415.881   &  0.074 &    -73.539  &    0.096 &   2    &          &         &      \\
IRS1 &  3  &   35  &   5.01 & 0.21   & 0.226  &  0.011  &    52.644   &  0.031 &     85.721  &    0.038 &   2    &          &         &  19  \\
IRS1 &  3  &   36  &   4.96 & 0.21   & 0.117  &  0.011  &    53.210   &  0.100 &     87.473  &    0.091 &   2    &          &         &      \\
IRS1 &  3  &   37  &   4.87 & $<$0.01  & 2.436  &  0.015  &  -413.873   &  0.039 &    -79.125  &    0.064 &   6    &   1.128  &  0.008  &      \\
IRS1 &  3  &   38  &   4.24 & 0.21   & 1.151  &  0.009  &  -413.040   &  0.017 &    -79.301  &    0.062 &   3    &   0.894  &  0.011  &      \\
IRS1 &  3  &   39  &   3.00 & 0.21   & 0.662  &  0.025  &  -393.282   &  0.088 &    -88.356  &    0.080 &   3    &   0.729  &  0.028  &      \\
IRS1 &  3  &   40  &   2.90 & 0.05   & 3.604  &  0.337  &  -410.419   &  0.182 &    -80.399  &    0.093 &   5    &   1.240  &  0.126  &      \\
IRS1 &  3  &   41  &   2.55 & 0.04   & 6.105  &  0.504  &  -411.855   &  0.223 &    -80.012  &    0.224 &   7    &   1.041  &  0.083  &      \\
IRS1 &  3  &   42  &   2.37 & 0.21   & 1.208  &  0.011  &  -409.351   &  0.267 &    -80.788  &    0.195 &   2    &          &         &  20  \\
IRS1 &  3  &   43  &  -0.41 & 0.04   & 0.273  &  0.014  &  -396.603   &  0.056 &    -87.986  &    0.066 &   6    &   1.438  &  0.091  &      \\
IRS1 &  3  &   44  &  -0.69 & 0.21   & 0.100  &  0.016  &  -395.770   &  0.178 &    -88.495  &    0.057 &   3    &          &         &      \\
IRS1 &  3  &   45  &  -1.12 & 0.06   & 0.226  &  0.014  &    20.329   &  0.126 &     47.082  &    0.116 &   4    &   1.318  &  0.150  &  21  \\
     &	   &	   &	    & 		    &	          & 	 & 	       &        & 	      & 	 & 	  & 	     & 	       &      \\
IRS1 &  4  &   0   &  26.74 & 0.21   & 0.885  &  0.044  &   311.963   &  0.018 &     70.049  &    0.047 &   3    &   1.302  &  0.024  &   0  \\
IRS1 &  4  &   1   &  18.66 & 0.21   & 1.220  &  0.013  &     4.095   &  0.033 &    -11.240  &    0.040 &   3    &   0.655  &  0.004  &   1  \\
IRS1 &  4  &   2   &  14.27 & 0.21   & 0.720  &  0.016  &   248.139   &  0.018 &     76.336  &    0.019 &   2    &          &         &   2  \\
IRS1 &  4  &   3   &  13.59 & 0.21   & 0.234  &  0.024  &     1.038   &  0.062 &     -1.994  &    0.067 &   1    &          & 	       &      \\
IRS1 &  4  &   4   &  13.06 & 0.01   & 4.779  &  0.180  &  -109.901   &  0.392 &   -122.328  &    0.130 &   5    &   0.758  &  0.033  &      \\
IRS1 &  4  &   5   &  13.02 & 0.21   & 0.788  &  0.016  &  -113.622   &  0.110 &   -123.445  &    0.018 &   2    &          &         &   4  \\
IRS1 &  4  &   6   &  12.68 & 0.21   & 0.924  &  0.013  &    81.655   &  0.021 &   -113.748  &    0.039 &   3    &   0.685  &  0.006  &   6  \\
IRS1 &  4  &   7   &  12.60 & 0.02   & 7.291  &  0.258  &     1.854   &  0.096 &      0.793  &    0.092 &   7    &   0.937  &  0.039  &      \\
IRS1 &  4  &   8   &  12.58 & 0.05   & 0.856  &  0.076  &   -85.740   &  0.044 &   -100.298  &    0.372 &   4    &   0.982  &  0.108  &   5  \\
IRS1 &  4  &   9   &  12.52 & 0.21   & 0.435  &  0.016  &    80.177   &  0.128 &   -113.481  &    0.059 &   2    &          &         &   7  \\
IRS1 &  4  &   10  &  11.75 & 0.21   & 0.288  &  0.016  &     0.580   &  0.313 &     -0.837  &    0.121 &   2    &          &         &   8  \\
IRS1 &  4  &   11  &  10.44 & 0.02   & 0.663  &  0.015  &    -0.788   &  0.178 &     -0.638  &    0.119 &   4    &   1.191  &  0.039  &   9  \\
IRS1 &  4  &   12  &  10.03 & 0.21   & 0.427  &  0.016  &  -242.408   &  0.041 &    -51.576  &    0.054 &   2    &          &         &  10  \\
IRS1 &  4  &   13  &   9.38 & 0.21   & 0.149  &  0.024  &    -1.933   &  0.097 &     17.095  &    0.105 &   1    &          & 	       &      \\
IRS1 &  4  &   14  &   9.26 & 0.21   & 2.295  &  0.064  &  -281.148   &  0.040 &    -96.016  &    0.030 &   3    &   0.603  &  0.020  &  11  \\
IRS1 &  4  &   15  &   8.96 & 0.21   & 0.171  &  0.024  &  -282.837   &  0.085 &   -103.703  &    0.108 &   1    &          & 	       &      \\
IRS1 &  4  &   16  &   8.84 & $<$0.01  & 3.742  &  0.012  &  -281.074   &  0.178 &    -91.360  &    0.282 &   4    &   0.738  &  0.002  &      \\
IRS1 &  4  &   17  &   8.74 & 0.21   & 0.298  &  0.016  &  -278.322   &  0.150 &   -104.772  &    0.045 &   2    &          &         &  13  \\
IRS1 &  4  &   18  &   8.67 & 0.02   & 0.730  &  0.025  &  -279.869   &  0.284 &    -88.895  &    0.363 &   4    &   0.931  &  0.039  &  14  \\
IRS1 &  4  &   19  &   8.64 & 0.21   & 0.335  &  0.024  &  -281.632   &  0.104 &    -73.758  &    0.342 &   3    &          &         &      \\
IRS1 &  4  &   20  &   8.53 & 0.21   & 0.179  &  0.024  &  -276.703   &  0.081 &   -120.125  &    0.102 &   1    &          & 	       &      \\
IRS1 &  4  &   21  &   8.12 & 0.02   & 1.318  &  0.067  &  -341.001   &  0.066 &    -56.718  &    0.047 &   4    &   0.894  &  0.057  &  16  \\
IRS1 &  4  &   22  &   7.48 & 0.21   & 0.734  &  0.016  &  -274.182   &  0.121 &   -105.905  &    0.114 &   2    &          &         &  17  \\
IRS1 &  4  &   23  &   6.64 & 0.21   & 0.277  &  0.016  &  -144.169   &  0.450 &   -135.609  &    0.060 &   2    &          &         &  18  \\
IRS1 &  4  &   24  &   5.85 & 0.21   & 0.198  &  0.016  &  -145.361   &  0.322 &   -135.708  &    0.087 &   2    &          &         &      \\
IRS1 &  4  &   25  &   4.96 & 0.21   & 0.180  &  0.016  &    53.398   &  0.089 &     87.300  &    0.138 &   2    &          &         &      \\
IRS1 &  4  &   26  &   4.62 & 0.05   & 0.300  &  0.017  &  -414.521   &  0.090 &    -79.880  &    0.147 &   5    &   1.557  &  0.126  &      \\
IRS1 &  4  &   27  &   4.61 & 0.21   & 0.327  &  0.016  &    92.086   &  0.043 &    -35.756  &    0.117 &   2    &          &         &      \\
IRS1 &  4  &   28  &   3.81 & 0.01   & 0.881  &  0.012  &  -394.200   &  0.138 &    -89.204  &    0.015 &   4    &   0.989  &  0.006  &      \\
IRS1 &  4  &   29  &   3.02 & 0.01   & 1.945  &  0.039  &  -412.984   &  0.089 &    -80.343  &    0.018 &   5    &   1.120  &  0.027  &      \\
IRS1 &  4  &   30  &   2.03 & 0.01   & 7.076  &  0.133  &  -410.001   &  0.160 &    -81.259  &    0.063 &   6    &   0.932  &  0.019  &  20  \\
IRS1 &  4  &   31  &  -0.30 & 0.06   & 0.360  &  0.017  &  -396.497   &  0.240 &    -88.936  &    0.248 &   6    &   1.814  &  0.131  &      \\
\enddata

\tablenotetext{a}{Half Power Full Widths (HPFW) are quoted for features with 
line profiles fit to a Gaussian model.}

\tablenotetext{b}{For features with measured proper motions, entries indicate
the identification numbers of motions, as enumerated in Table\,3.}
\end{deluxetable}

\begin{deluxetable}{cccrrrlrlcccc}
\tablewidth{0pt}
\rotate
\tabletypesize{\scriptsize}
\tablecaption{Water Maser Proper Motions in NGC2071\,IRS1 \& IRS3}
\tablehead{
\colhead{Object} & \colhead{Motion} & \colhead{V$_{\rm LOS}$} & 
\colhead{E-W} & \colhead{N-S} & \colhead{E-W} & 
\colhead{Unc.} & \colhead{N-S} & \colhead{Unc.} & 
\multicolumn{4}{c}{Corresponding Features\tablenotemark{b}} \\

\colhead{} & \colhead{No.} & \colhead{} & 
\colhead{Offset} & \colhead{Offset} & \colhead{Vel.\tablenotemark{a}} & 
\colhead{} & \colhead{Vel.\tablenotemark{a}} & \colhead{} & 
\colhead{1} & \colhead{2} & \colhead{3} & \colhead{4}  \\

\colhead{} & \colhead{} & \colhead{(\kms)} & 
\colhead{(mas)} & \colhead{(mas)} & \colhead{(\kms)} & 
\colhead{(\kms)} & \colhead{(\kms)} & \colhead{(\kms)} & 
\colhead{} & \colhead{} & \colhead{} & \colhead{} \\        
}

\startdata
IRS3  & 0   &     24.86    &   -166.310 &   -97.542 &   -22.35  &    0.28  &  -35.03  &    0.94  &       &    1 &    0  &   1 \\
IRS3  & 1   &     15.50    &     84.444 &    -2.506 &    18.86  &    0.14  &  -10.62  &    0.03  &     2 &    4 &    7  &     \\
IRS3  & 2   &     13.36    &    108.188 &  -234.607 &    16.91  &    0.17  &  -16.67  &    0.04  &       &    6 &   12  &   4 \\
IRS3  & 3   &     11.64    &      6.196 &    40.722 &    25.46  &    9.22  &   -0.49  &    0.85  &     7 &    9 &   13  &     \\
IRS3  & 4   &      4.83    &      1.096 &     2.123 &    -7.94  &    0.20  &    2.21  &    0.07  &       &   13 &   21  &   8 \\
IRS3  & 5   &      4.31    &      3.037 &     1.305 &     1.22  &    0.53  &   -2.01  &    0.03  &       &   16 &   25  &  11 \\
IRS3  & 6\tablenotemark{c}   &      3.15    &     -7.509 &     7.701 &     0.00  &    0.00  &    0.00  &    0.00  &    18 &   19 &   28  &  13 \\
IRS3  & 7   &      2.73    &     -8.637 &     9.948 &    -4.30  &    0.47  &    6.90  &    0.07  &    21 &      &   30  &  14 \\
      &     & 		   & 	        & 	    & 	        & 	   & 	      & 	 & 	 &      &       &     \\
IRS1  & 0   &     26.82    &    311.503 &    69.055 &    12.02  &    0.50  &    16.84 &    0.52  &       &    1 &    1  &   0 \\
IRS1  & 1   &     18.44    &      4.250 &   -11.030 &    -2.23  &    0.37  &   -11.02 &    0.53  &       &    4 &    2  &   1 \\
IRS1  & 2   &     14.27    &    247.208 &    76.341 &    22.97  &    0.03  &    -6.05 &    0.04  &       &   12 &    8  &   2 \\
IRS1  & 3   &     13.83    &    248.715 &    75.406 &    22.20  &    0.01  &    -6.15 &    0.50  &    20 &   13 &    9  &     \\
IRS1  & 4   &     13.03    &   -113.669 &  -123.082 &     3.38  &    0.16  &   -14.42 &    0.22  &       &   15 &   11  &   5 \\
IRS1  & 5   &     12.55    &    -85.972 &  -100.361 &     5.95  &    1.79  &   -13.13 &    0.36  &    21 &   17 &       &   8 \\
IRS1  & 6   &     12.50    &     82.216 &  -113.766 &   -11.84  &    1.01  &    -5.51 &    0.07  &       &   18 &   15  &   6 \\
IRS1  & 7   &     12.19    &     80.813 &  -113.115 &    -9.11  &    1.04  &   -12.21 &    1.24  &    25 &      &   16  &   9 \\
IRS1  & 8   &     11.98    &      1.392 &     0.450 &    -8.36  &    0.82  &   -27.09 &    2.14  &    24 &   19 &   17  &  10 \\
IRS1  & 9   &     10.43    &     -0.881 &    -0.495 &     2.29  &    0.47  &    -7.14 &    2.24  &       &   22 &   18  &  11 \\
IRS1  & 10  &     10.21    &   -241.748 &   -51.511 &   -15.89  &    2.10  &    -6.71 &    0.26  &    28 &      &   20  &  12 \\
IRS1  & 11  &      9.24    &   -281.153 &   -96.119 &     1.36  &    0.01  &    -3.48 &    0.05  &       &   24 &   22  &  14 \\
IRS1  & 12\tablenotemark{d}  &      8.76    &   -281.020 &   -91.615 &    -0.68  &    0.00  &     3.21 &    0.00  &    34 &   28 &   25  &  16 \\
IRS1  & 13  &      8.73    &   -278.479 &  -104.944 &     2.18  &    0.14  &    -2.11 &    0.53  &       &   26 &   24  &  17 \\
IRS1  & 14  &      8.71    &   -279.695 &   -88.523 &    -1.80  &    0.11  &    -6.26 &    0.40  &    32 &   29 &   26  &  18 \\
IRS1  & 15  &      8.66    &   -335.146 &   -55.301 &   -22.07  &    0.20  &    -7.74 &    2.13  &    35 &   27 &   23  &     \\
IRS1  & 16  &      8.13    &   -340.311 &   -56.521 &   -13.84  &    1.77  &    -9.91 &    0.20  &       &   31 &   30  &  21 \\
IRS1  & 17  &      7.56    &   -274.396 &  -106.040 &     5.88  &    0.28  &    -1.80 &    1.51  &       &   33 &   32  &  22 \\
IRS1  & 18  &      6.57    &   -144.558 &  -135.460 &    10.03  &    0.10  &    -8.58 &    0.09  &       &   35 &   33  &  23 \\
IRS1  & 19  &      4.98    &     52.629 &    85.737 &     3.51  &    0.15  &   -11.94 &    0.31  &    42 &   37 &   35  &     \\
IRS1  & 20  &      2.20    &   -409.432 &   -80.692 &   -11.64  &    0.53  &   -19.31 &    0.22  &       &   41 &   42  &  30 \\
IRS1  & 21  &     -1.22    &     20.316 &    47.117 &    11.43  &    0.17  &    -1.46 &    0.02  &    48 &   42 &   45  &     \\
\enddata
\tablenotetext{a}{Measured proper motions expressed as a velocity, 
assuming a distance of 390 pc.}
\tablenotetext{b}{Identification numbers of maser features at each epoch (as 
enumerated in Table\,2) that contribute to the listed proper motions.}
\tablenotetext{c}{IRS\,3 motion 6 is zero because the corresponding
  feature was used to align the maps of the four epochs.}
\tablenotetext{d}{IRS\,1 motion 12 corresponds to the maser feature
  used to align the maps of the four epochs.  An offset is added to
  all the motions so that the net apparent motion of the IRS1 region
  is zero. Motion 12 corresponds to that offset.}
\end{deluxetable}

\clearpage

\begin{figure}
\plotone{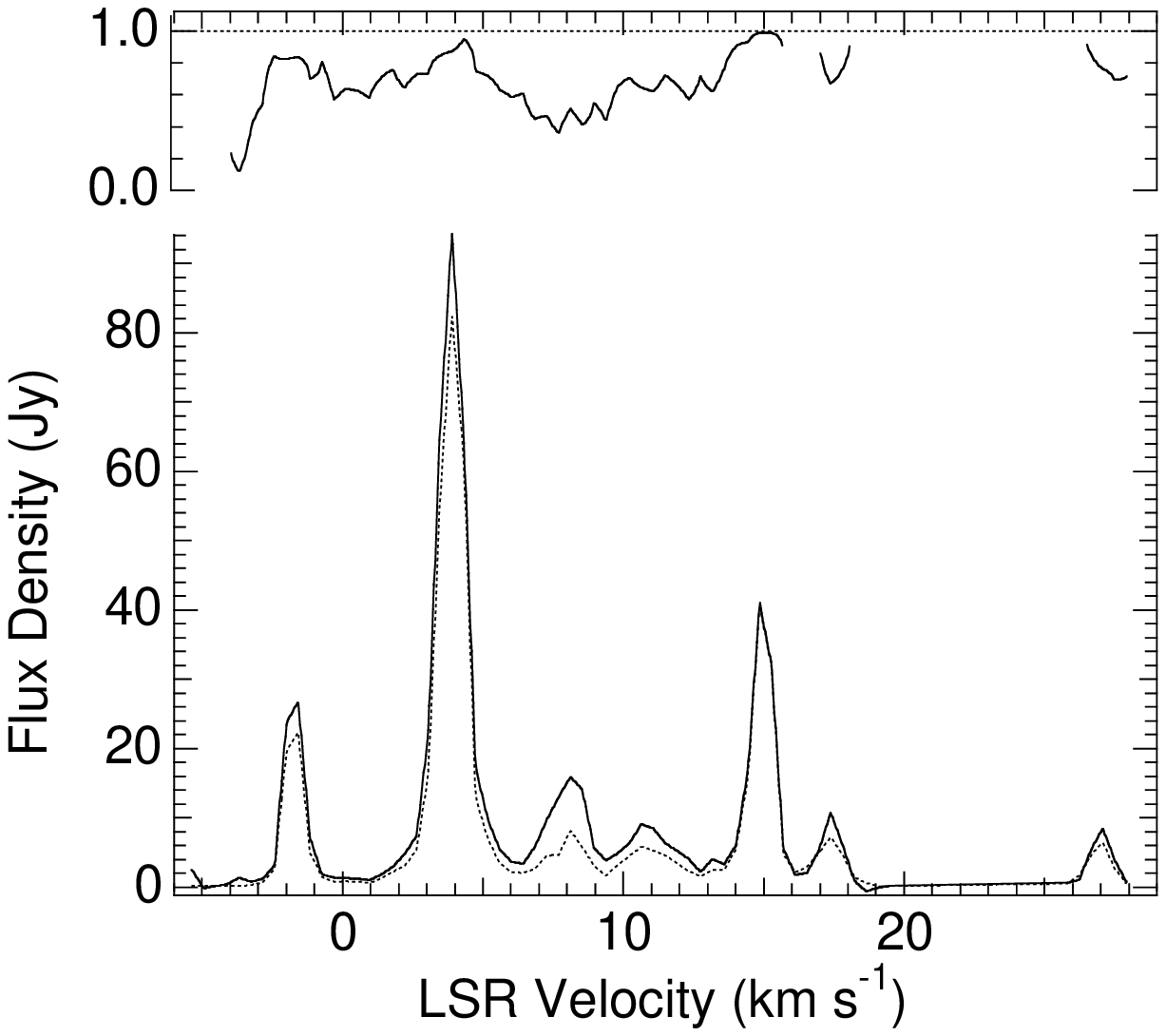}
\end{figure}

\newpage

\begin{figure}
\epsscale{0.6}
\plotone{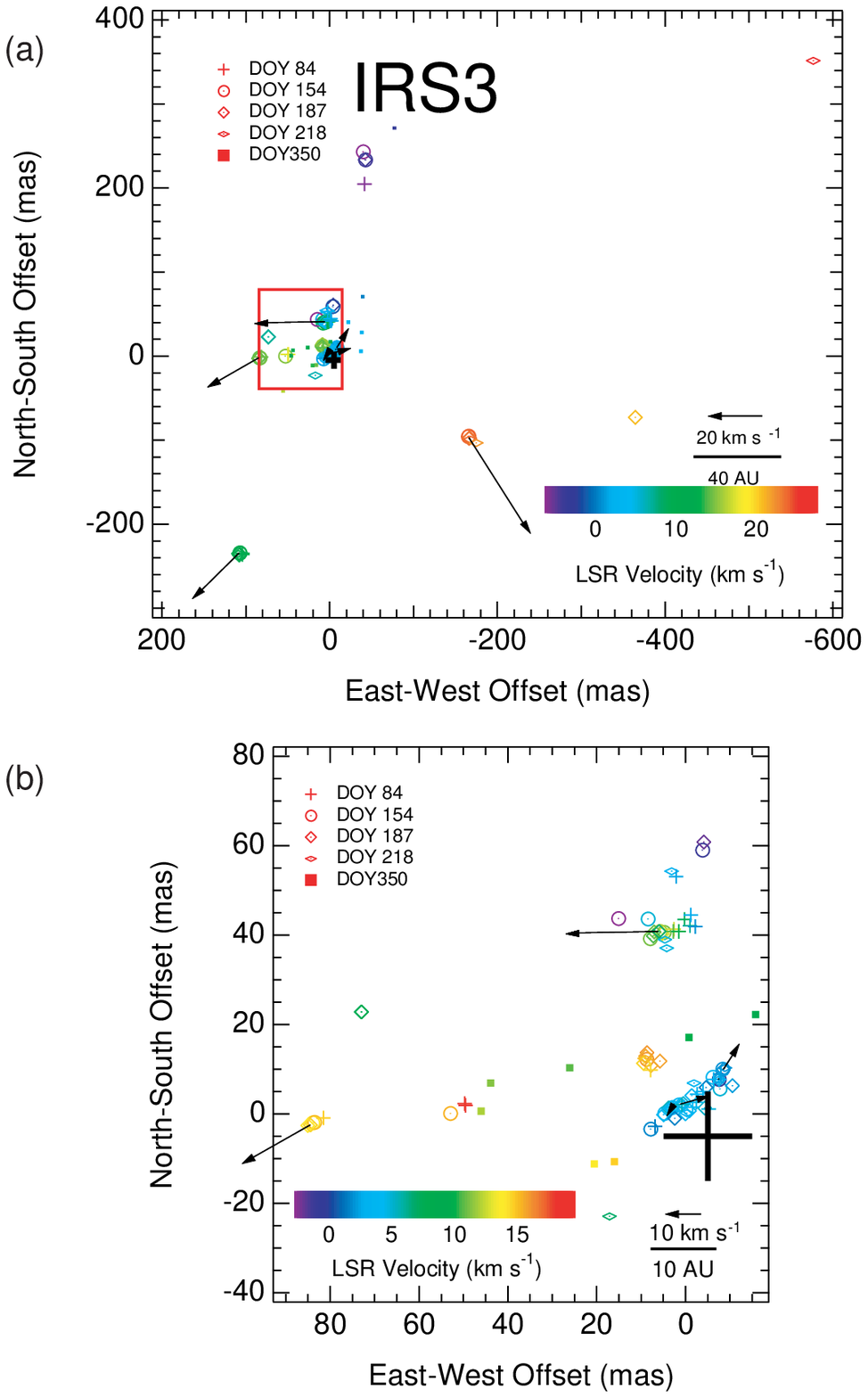}
\end{figure}

\newpage

\begin{figure}
\epsscale{0.6}
\plotone{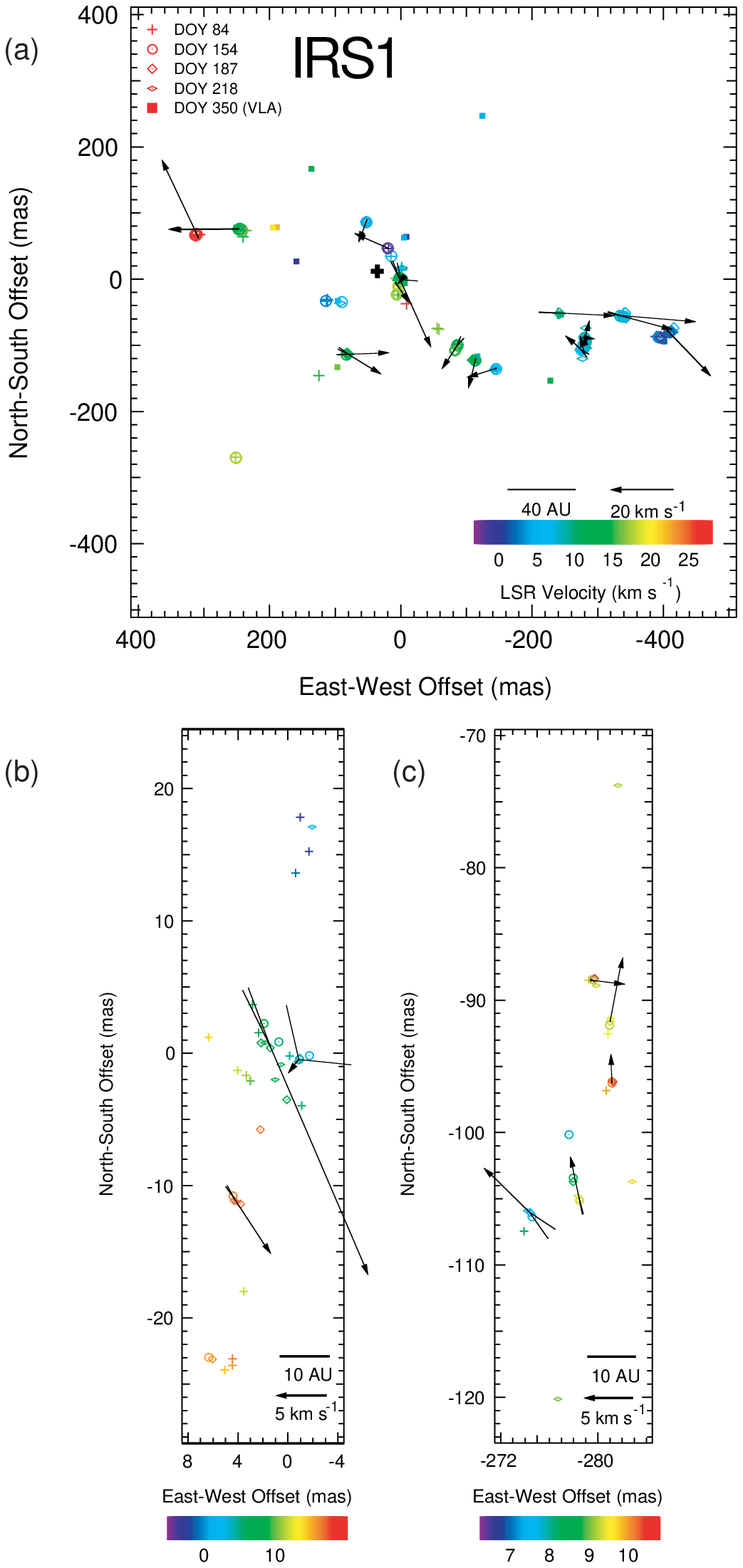}
\end{figure}

\newpage

\begin{figure}
\epsscale{0.6}
\plotone{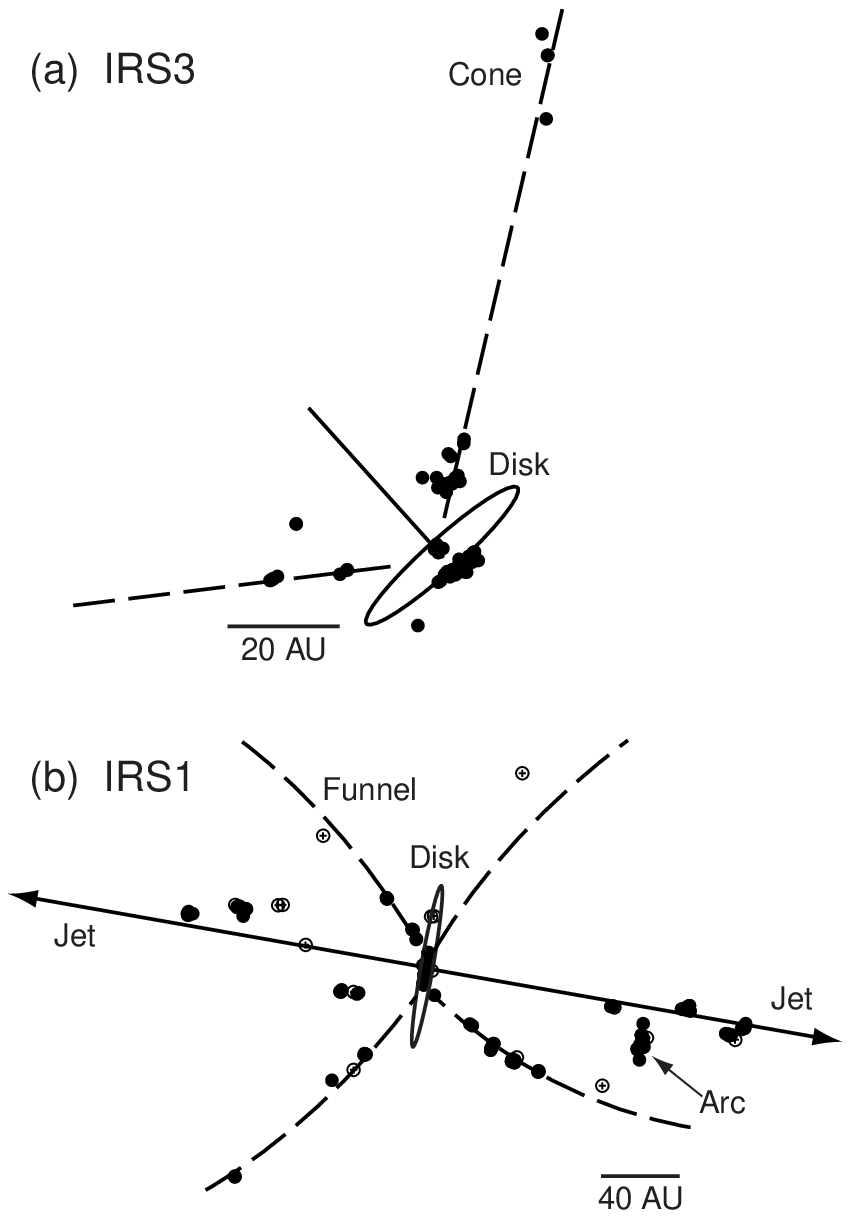}
\end{figure}

\end{document}